\documentclass[twocolumn,superscriptaddress,amsmath,amssymb,aps,nofootinbib,prx]{revtex4-2}
\usepackage{amsmath,amssymb}
\usepackage{graphicx}
\usepackage{dcolumn}
\usepackage{bm}
\usepackage{multirow}
\usepackage{textcomp}
\usepackage{url}
\usepackage{enumitem}
\usepackage{float}
\usepackage[dvipsnames]{xcolor}
\usepackage[colorlinks=true, allcolors=cyan]{hyperref}
\usepackage{orcidlink}
\newcommand{\bra}[1]{\langle{#1}|}
\newcommand{\ket}[1]{|{#1}\rangle}

\begin{document}

\title{Reduced Density Matrices from Renormalization-Group Flows}

\author{X. Zhen\,\orcidlink{0009-0000-1806-4123}}
\affiliation{School of Physics, and State Key Laboratory of Nuclear Physics and Technology, Peking University, Beijing 100871, China}
\author{Z.~Y.~Meng\,\orcidlink{0009-0007-2538-7706}}
\affiliation{School of Physics, and State Key Laboratory of Nuclear Physics and Technology, Peking University, Beijing 100871, China}
\author{J. C. Pei\,\orcidlink{0000-0002-9286-1304}}\email[]{peij@pku.edu.cn}
\affiliation{School of Physics, and State Key Laboratory of Nuclear Physics and Technology, Peking University, Beijing 100871, China}
\affiliation{Southern Center for Nuclear-Science Theory (SCNT), Institute of Modern Physics, Chinese Academy of Sciences, Huizhou 516000, China}
\author{F. R. Xu\,\orcidlink{0000-0001-6699-0965}}\email[]{frxu@pku.edu.cn}
\affiliation{School of Physics, and State Key Laboratory of Nuclear Physics and Technology, Peking University, Beijing 100871, China}
\affiliation{Southern Center for Nuclear-Science Theory (SCNT), Institute of Modern Physics, Chinese Academy of Sciences, Huizhou 516000, China}

\date{\today}

\begin{abstract}
We reconstruct complete one- and two-body reduced density matrices (RDMs) by a single collective backward evolution along the in-medium similarity renormalization-group flow.
Benchmarks show improved accuracy and physical consistency over conventional elementwise evolution.
In neutron matter, two-body RDMs reveal spin-triplet dominance in interaction-dependent Fermi-sea depletion trends.
Meanwhile, the forward--backward energy difference provides an internal diagnostic of many-body truncation errors, whose quantification remains challenging in nuclear \textit{ab initio} calculations.
\end{abstract}

\maketitle

\textit{Introduction.---}
\textit{Ab initio} methods seek systematically improvable solutions of the many-body Schrödinger equation based on realistic nuclear interactions~\cite{Machleidt2024PPNP, Machleidt2023FBS, Ekstrom2023FP}. 
Scalable approaches such as coupled-cluster theory~\cite{Coester1958NP,Coester1960NP,Dean2004PRC,Hagen2014RPP,Shavitt2009CUP} and the in-medium similarity renormalization group~\cite{Tsukiyama2011PRL,Morris2015PRC,Hergert2016PhysRep} (IMSRG) avoid full wave-function construction, enabling calculations across a broad range of the nuclear chart~\cite{Stroberg2019ARNPS,Hergert2020FP,Stroberg2017PRL,Morris2018PRL,Gysbers2019NatPhys,Stroberg2021PRL,Miyagi2022PRC,Hu2022NatPhys}. 
However, two non-negligible limitations remain.
First, evaluating observables beyond the energy generally requires additional, observable-specific calculations, making systematic studies of many observables computationally demanding~\cite{Hagen2014RPP,Bacca2013PRL,Sobczyk2025PRL,Morris2015PRC,Schuster2014PRC,Parzuchowski2017PRC,Miyagi2024PRL}.
Second, many-body truncation errors are difficult to quantify because exact benchmarks are restricted to the smallest systems, whereas other calculations at higher truncation ranks rapidly become computationally prohibitive~\cite{Navratil2009JPG, Svensson2026PRC, Drischler2026PRC, Li2026arXiv, Binder2013PRC, Hagen2007PRC, Heinz2021PRC, Stroberg2024PRC, Raimondi2018PRC, Barbieri2022PRC, Marino2024PRC, Marino2026PRC}.
Recent advances in full configuration-interaction quantum Monte Carlo~\cite{Hu2026PRL, Hu2026arXiv1, Hu2026arXiv2, Booth2009JCP, Overy2014JCP} (FCIQMC) offer valuable benchmarks, but systematic large-scale comparisons remain computationally demanding.
A reusable representation of correlated states, together with an internal uncertainty diagnostic, would help address both limitations.

Reduced density matrices (RDMs) provide such a reusable representation.
Complete one- and two-body RDMs determine all expectation values up to two-body rank through direct contractions and expose correlations including natural occupations and pair structure, beyond the information contained in the energy alone~\cite{Lowdin1955PR, Coleman1963RMP, Overy2014JCP, DePrince2024WCMS}.
Because of the computational cost, most scalable methods provide selected observables or partial information about RDMs, while complete low-rank RDMs are not routinely available in nuclear \textit{ab initio} calculations~\cite{Soma2014PRC, Tichai2019PRC, Hoppe2021PRC, Burrows2018PRC, Hagen2014RPP, Morris2015PRC}. 
For methods based on continuous unitary transformations, backward state evolution provides a route to reconstructing these RDMs and is dual to forward operator evolution in exact theory.
Under truncated approximations, the forward--backward path-consistency defect can serve as an internal diagnostic of the approximation.

Building on this duality, in this Letter we formulate inverse state evolution in reduced-density space by closing the Bogoliubov--Born--Green--Kirkwood--Yvon (BBGKY) hierarchy~\cite{Born1946PRSA} at the two-body level, using different closures to approximate the three-body cumulant.
The underlying forward flow is generated using the IMSRG truncated at the normal-ordered two-body level [IMSRG(2)].
A single collective backward evolution thus provides complete one- and two-body RDMs for subsequent calculations of observables while retaining the computational scaling of IMSRG(2).
The forward--backward energy discrepancy offers a route to diagnosing and quantifying many-body truncation uncertainties.
Benchmarks across diverse toy models and chiral nuclear matter demonstrate improved reconstruction accuracy and physical consistency relative to conventional elementwise evolution, while supporting the proposed uncertainty diagnostic.

\begin{figure*}[t]
    \centering
    \includegraphics[width=0.9\linewidth]{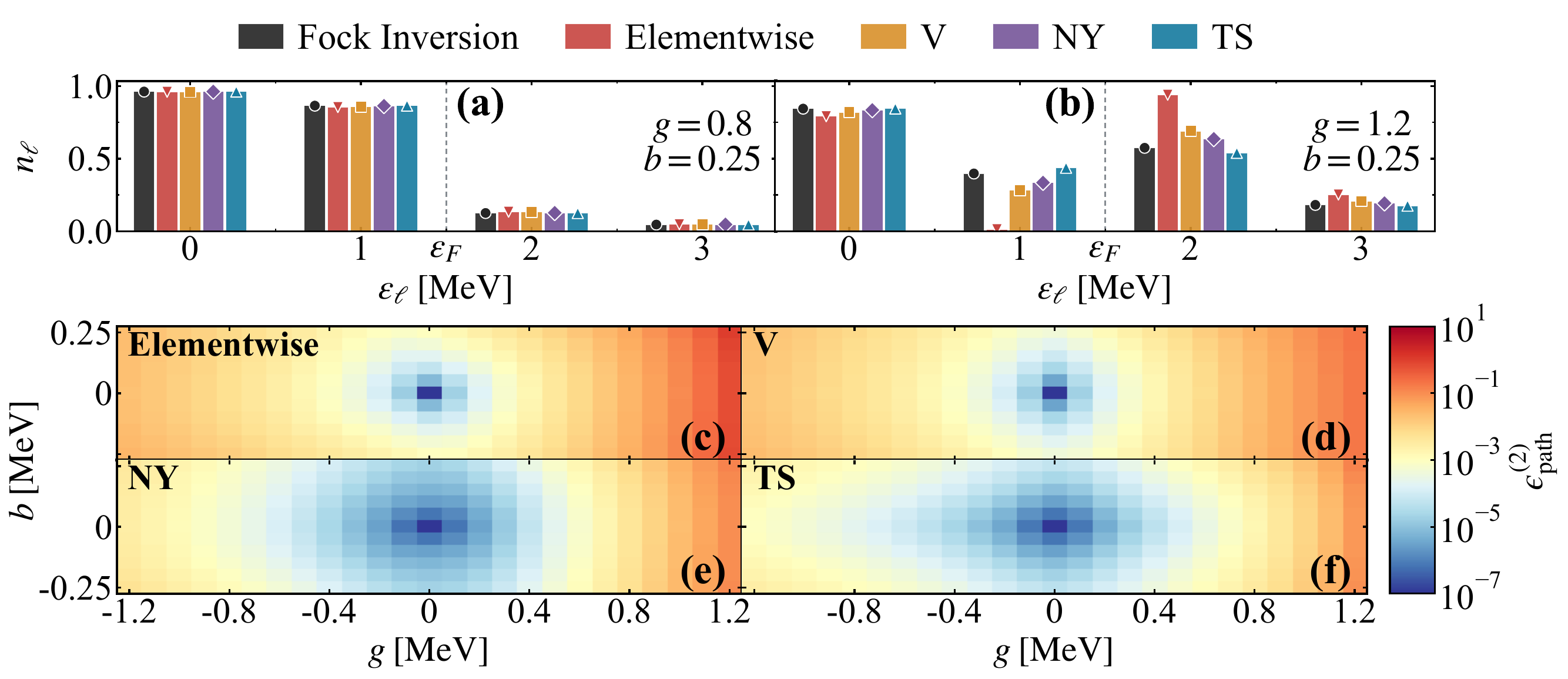}
\caption{\label{fig:second}
Reduced density matrices calculated for the $A=4$ pairing-plus-particle-hole (P3H) model.
The occupations $n_\ell$ in panels (a) and (b) correspond to $(g,b)=(0.8,0.25)$ and $(1.2,0.25)\,\mathrm{MeV}$, respectively.
Bars compare Fock-space inversion, elementwise evolution, and BBGKY(2) with the V, NY, and TS closures.
The dashed line marks the Fermi energy $\varepsilon_F=1.5\,\mathrm{MeV}$.
The lower panels show the mixed path-consistency error $\epsilon_{\rm path}^{(2)}$, which is defined as the relative Frobenius distance between each reconstructed 2-RDM and the corresponding Fock-space-inversion result.
All calculations are averaged over the three generators.}
\end{figure*}

\textit{Method.---}
The IMSRG uses a continuous unitary transformation $U(s)$ to decouple the off-diagonal part of Hamiltonian $H(0)$,
\begin{align}
    H(s)&=U(s)H(0)U^\dagger(s),\
    \eta(s)\equiv\frac{{\rm d}U(s)}{{\rm d}s}U^\dagger(s),\nonumber\\
&\qquad \quad\frac{{\rm d}H(s)}{{\rm d}s}=[\eta(s),H(s)], 
\end{align}
where $s$ is the flow parameter and $\eta(s)$ is the anti-Hermitian generator~\cite{Hergert2016PhysRep, Tsukiyama2011PRL}.
All flowing operators are normal-ordered with respect to the reference state $\ket{\Phi}$ and truncated at the two-body level, which defines the IMSRG(2) approximation used here. For recent applications to nuclear matter, see Refs.~\cite{Cook2026arXiv,Udiani2024MSU,Yu2026MSU,Cook2026MSU,Zhen2025PLB}.
At the endpoint $s_f$, the stored trajectory defines an approximate correlated state $\ket{\Psi^\prime}$ and its density operator,
\begin{align}
    \ket{\Psi'}&=U^\dagger(s_f)\ket{\Phi},\ \hat\varrho_\Phi\equiv\ket{\Phi}\bra{\Phi},\nonumber\\
    \hat\varrho_{\Psi'}
    &\equiv\ket{\Psi'}\bra{\Psi'}
    =U^\dagger(s_f)\hat\varrho_\Phi U(s_f).\label{eq:reverse}
\end{align}
Because the operator flow is truncated, $\ket{\Psi'}$ generally differs from the exact ground state $\ket{\Psi}$. 
Our targets are complete one- and two-body RDMs
\begin{align}
    \varrho^{(1)}_{pq}
    &=\bra{\Psi'}a_p^\dagger a_q\ket{\Psi'},
    &
    \varrho^{(2)}_{pqrs}
    &=\bra{\Psi'}a_p^\dagger a_q^\dagger a_s a_r\ket{\Psi'}.
    \label{eq:rdm}
\end{align}
Direct Fock-space inversion as in Eq.~(\ref{eq:reverse}) can track the IMSRG(2) trajectory exactly backward, which is combinatorially expensive.
Alternatively, evolving all 2-RDM elements separately requires $\sim\mathcal O(N_{\rm sp}^{10})$ operations and introduces an independent operator truncation for each element~\cite{Hergert2016PhysRep, Morris2015PRC, Tsukiyama2011PRL}, where $N_{\rm sp}$ is the number of single-particle orbitals. 
In contrast to these two routes, we propagate all RDM elements collectively by replaying the stored generator trajectory backward. 

\begin{figure*}[t]
    \centering
    \includegraphics[width=1.0\linewidth]{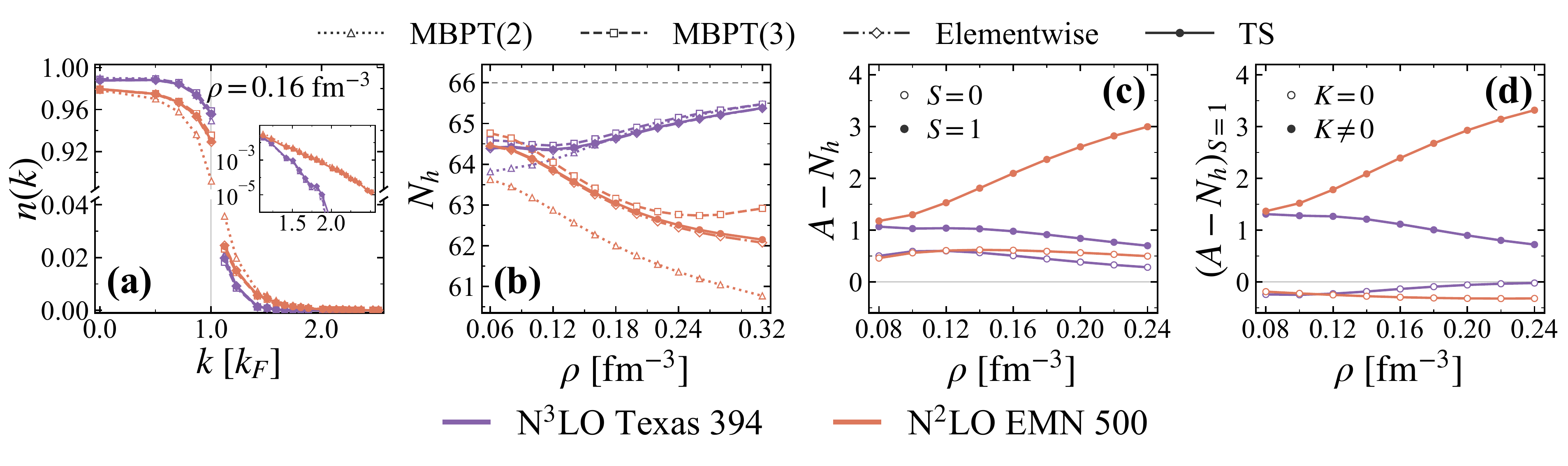}
    \caption{\label{fig:third}
Momentum distributions and two-body cumulant decomposition of pure neutron matter. Purple and orange denote the N$^3$LO Texas and N$^2$LO EMN interactions, respectively. (a) Momentum distributions at $\rho=0.16~\mathrm{fm}^{-3}$ from MBPT(2), MBPT(3), elementwise IMSRG(2), and IMSRG(2)-BBGKY(2) with the TS closure. The inset shows the high-momentum tails on a logarithmic vertical scale. (b) Density dependence of $N_h$ [Eq.~(10)] for the same methods. (c) Spin-singlet (open symbols) and spin-triplet (filled symbols) contributions to the depletion $A-N_h$, summed over all pair center-of-mass momenta. (d) Contributions from $\bm{K}=0$ (open symbols) and $\bm{K}\neq0$ (filled symbols) to the triplet depletion $(A-N_h)_{S=1}$.}
\end{figure*}

Because RDMs are vacuum-normal-ordered, the one- and two-body parts of the reference-normal-ordered generator are first converted as
\begin{align}
    \widetilde{\eta}^{(1)}_{pq}
    &=\eta^{(1)}_{pq}-\sum_i n_i\eta^{(2)}_{piqi},
    &
    \widetilde{\eta}^{(2)}&=\eta^{(2)},
    \label{eq:eta_restore}
\end{align}
where $n_i$ denote the reference occupation numbers. 
Starting from $\hat\varrho(s_f)=\hat\varrho_\Phi$, we integrate
\begin{align}
    \frac{{\rm d}\hat\varrho(s)}{{\rm d}s}
    =[\widetilde\eta(s),\hat\varrho(s)]
\end{align}
from $s_f$ back to zero. Its one- and two-body reductions are
\begin{align}
    \frac{{\rm d}\varrho^{(1)}}{{\rm d}s}
    &= [\widetilde\eta^{(1)},\varrho^{(1)}]
    +\operatorname{Tr}_2[\widetilde\eta^{(2)},\varrho^{(2)}],\\
    \frac{{\rm d}\varrho^{(2)}}{{\rm d}s}
    &= [\widetilde\eta^{(1)}_1+\widetilde\eta^{(1)}_2
       +\widetilde\eta^{(2)}_{12},\varrho^{(2)}]
    +\operatorname{Tr}_3[\widetilde\eta^{(2)}_{13}
       +\widetilde\eta^{(2)}_{23},\varrho^{(3)}].
\end{align}
The subscripts label particle slots rather than indices. These are the first two equations of the BBGKY hierarchy~\cite{Born1946PRSA, Bogoliubov1946JPhysUSSR, Kirkwood1946JCP, Yvon1935Hermann} and require a closure approximation for the three-body RDM. We use the fermionic cumulant expansion~\cite{Akbari2012PRB, Lackner2015PRA}
\begin{align}
    \lambda^{(2)}
    &=\varrho^{(2)}-2\varrho^{(1)}\wedge\varrho^{(1)},\nonumber\\
    \varrho^{(3)}
    &=6\varrho^{(1)}\wedge\varrho^{(1)}\wedge\varrho^{(1)}
    +9\lambda^{(2)}\wedge\varrho^{(1)}+\lambda^{(3)},
    \label{eq:closure}
\end{align}
where $\wedge$ denotes the antisymmetrized tensor product and $\lambda^{(k)}$ the irreducible $k$-body cumulant.
At the BBGKY(2) level, $\lambda^{(3)}$ is approximated as a functional of $\varrho^{(1)}$ and $\varrho^{(2)}$ using three closures considered in this work named Valdemoro~\cite{Colmenero1993PRA} (V), Nakatsuji--Yasuda~\cite{Nakatsuji1996PRL} (NY) and Tohyama--Schuck~\cite{Tohyama2014EPJA, Tohyama2017EPJA} (TS). 
When untruncated, the physical-time and inverse IMSRG evolutions are commutator flows along different trajectories on the same unitary orbit, motivating the use of these closures, which have been extensively benchmarked in time-dependent density-matrix theory~\cite{Lackner2015PRA,Lackner2017PRA,Donsa2023PRResearch,Tohyama2019EPJA}.
Reconstructing only the three-body elements required by the contractions retains the $\mathcal O(N_{\rm sp}^{6})$ scaling of IMSRG(2).
Also, to preserve contraction consistency, we additionally apply a minimum-norm projection to the BBGKY tangent at every integration step~\cite{Joost2022PRB, Benedikter2018AHP, Kronke2018PRA}.
All theoretical details can be found in Sec.~I of the Supplemental Material~\cite{SM}.

\textit{Results.---}
We first benchmark IMSRG(2)-BBGKY(2) in the pairing-plus-particle-hole (P3H) model~\cite{Smith2025PRC, HjorthJensen2010JPG}, with four nucleons in four doubly degenerate, equally spaced levels separated by $\delta=1.0\,\mathrm{MeV}$. The couplings $g$ and $b$ control pairing and pair-breaking excitations, respectively. The Hamiltonian and numerical setup are given in Sec.~II of the Supplemental Material~\cite{SM}. Exact inversion of each stored trajectory in the fixed-$A$ Fock space provides a reference that retains the underlying IMSRG(2) truncation error.

Fig.~\ref{fig:second}(a,b) compares diagonal occupations in the fixed single-particle energy basis. 
The points $(g,b)=(0.8,0.25)$ and $(1.2,0.25)\,\mathrm{MeV}$ respectively represent weak and strong couplings among 274 conditions. 
At the weakly coupled point in panel (a), all replays reproduce the Fock-inversion result with only small deviations. 
In panel (b), when the coupling becomes stronger, elementwise evolution substantially distorts the occupation transfer across the Fermi energy, whereas NY and TS closures remain much closer to the path-exact Fock-inversion result, and closure V gives an intermediate improvement. 
Panels (c--f) extend this comparison over the full $(g,b)$ plane using the mixed path-consistency defect
\begin{equation}
\epsilon_{\rm path}^{(2)}
=
\frac{
\left\|\varrho^{(2)}_X-
\varrho^{(2)}_{\mathrm{Fock}}\right\|_{\mathrm F}
}{
\left\|\varrho^{(2)}_{\mathrm{Fock}}\right\|_{\mathrm F}
},
\label{eq:path_defect}
\end{equation}
where $X$ denotes different reconstruction strategies, compared to Fock-space inversion. This quantity has been averaged over the White, arctangent, and imaginary-time generators after evaluating the norm, while the generator dependence has been found negligible~\cite{SM}. 
Among the three closures, NY and TS outperform elementwise evolution at all 274 nontrivial points, reducing the median defect by factors of approximately $12.8$ and $16.0$. 
Again, V generally provides a smaller improvement. 
Throughout all calculations, we monitor physical consistency during the reverse flow. 
For all replays, Hermiticity and contraction consistency are preserved to numerical precision, with residuals below $3\times10^{-14}$. 
We further assess the necessary $N$-representability, finding substantially smaller violations for the BBGKY reconstructions than for conventional elementwise evolution.
The diagnostics and complete flow-resolved results are provided in Secs.~II and~III of the Supplemental Material~\cite{SM}.

Next, we apply this framework to chiral neutron matter. We consider $A=66$ neutrons and $N_{\rm sp}=1030$ single-particle states in a cubic box with periodic boundary conditions, a commonly used setup~\cite{Hagen2014PRC,Jiang2020PRC,Jiang2024PRC,
      Marino2024PRC,Sobczyk2025PRL,Zhen2025PLB,
      Hu2026PRL,Marino2026PRC,Cook2026arXiv} for which finite-size effects are known to be relatively small.
For comparison, we employ two chiral nuclear interactions: the recently developed N$^3$LO Texas interaction~\cite{Hu2026PRL1} ($\Lambda=394~\mathrm{MeV}$), which has demonstrated good predictive performance, and the well-established N$^2$LO EMN interaction~\cite{Entem2017PRC,Drischler2019PRL} ($\Lambda=500~\mathrm{MeV}$), which has been extensively applied and benchmarked.
Both calculations include the normal-ordered two-body contributions from the corresponding nonlocal three-nucleon interactions.
Panel (a) of Fig.~\ref{fig:third} compares the momentum distributions for the two interactions at $\rho=0.16~\mathrm{fm}^{-3}$, obtained from the 1-RDMs calculated using MBPT(2), MBPT(3), elementwise IMSRG(2) evolution, and IMSRG(2)-BBGKY(2) replay with the TS closure for comparisons. 
All IMSRG calculations here and below use arctangent generators.
For the N$^2$LO EMN interaction, the MBPT(2) result deviates visibly, whereas all other calculations are mutually consistent. 
With results consistent with the replay, conventional elementwise evolution is approximately fifteen times less efficient to build the one-body RDMs (1-RDMs) under the same computational conditions.
The distinct perturbative behavior, together with the contrasting high-momentum tails shown in the inset of panel (a), clearly highlights the different correlation structures generated by the two interactions.
Panel (b) of Fig.~\ref{fig:third} shows the number of neutrons remaining in the reference Fermi sea,
\begin{align}
    N_h=\sum_{{\rm holes}}g(k)n(k),
\end{align}
where $g(k)$ is the degeneracy at momenta $k$. 
A smaller $N_h$ corresponds to a stronger depletion~\cite{Sammarruca2014PRC,Yang2019PRC,Rios2009PRC}. 
The EMN interaction produces a predominantly monotonic increase in depletion with density, while the Texas interaction exhibits a turnover. 
Although visible in the 1-RDM, this contrast cannot be resolved into contributions from different two-body correlation sectors using one-body information alone.
Panel (c) resolves the depletion $A-N_h$ into spin-singlet ($S=0$) and spin-triplet ($S=1$) contributions through decompositions of the two-body cumulant, summed over all pair center-of-mass momenta $\bm{K}$. 
The two interactions give comparable singlet contributions, while the triplet sector exhibits markedly different density dependence. 
For Texas, the triplet contribution decreases with density, whereas the singlet contribution initially increases before decreasing. 
Their sum therefore develops a maximum. 
For EMN, the increasing triplet contribution sustains the growth of the total depletion.
Panel (d) further separates the triplet contribution into $\bm{K}=0$ and $\bm{K}\neq0$ sectors. 
The substantial finite-$\bm{K}$ contribution shows that the integrated depletion cannot be understood from back-to-back pairs ($\bm{K}=0$) alone. 
Together, these decompositions locate the principal interaction-dependent contrast in the triplet sector and expose its pair-momentum structure. 
They illustrate how the reconstructed two-body RDMs (2-RDMs) connect a one-body signature to the underlying two-body correlations, beyond the information available from momentum distributions alone.

Finally, IMSRG(2)-BBGKY(2) offers a route to diagnosing the intrinsic many-body truncation error of IMSRG(2). 
Contracting the reconstructed RDMs with the initial Hamiltonian gives a second energy evaluation $E_{\rm BBGKY(2)}$, with the same Hamiltonian truncation as in the forward calculation. 
When the truncation is perturbatively controlled, the signed energy errors can be formally expanded in a dimensionless parameter $\varepsilon$ characterizing the omitted many-body contributions,
\begin{align}
d_{\rm MB}
\equiv
\frac{E_{\rm IMSRG(2)}-E_{\rm exact}}{A}
&=\sum_{i=p}^{\infty}m_i\varepsilon^i,\\
d_{\rm R}
\equiv
\frac{E_{\rm BBGKY(2)}-E_{\rm exact}}{A}
&=\sum_{j=q}^{\infty}n_j\varepsilon^j.
\end{align}
Here, $p$ and $q$ are the leading non-vanishing orders, and $m_i$ and $n_i$ are the expansion coefficients. 
As truncated dual descriptions of the same stored trajectory, they can be expected to share the leading omitted many-body structures. 
We therefore generically expect $p=q$, except when a leading coefficient vanishes through symmetry or accidental cancellation. 
The forward--backward discrepancy is then
\begin{align}
d_{\rm path}
\equiv d_{\rm R}-d_{\rm MB}
=\sum_{i=p}^{\infty}(n_i-m_i)\varepsilon^i.
\end{align}
Provided that $n_p-m_p\neq0$,
\begin{align}
d_{\rm MB}
=\frac{m_p}{n_p-m_p}\,d_{\rm path}
\left[1+\mathcal{O}(\varepsilon)\right].
\end{align}
At leading order, the replay discrepancy $d_{\rm path}$ thus tracks the scale of the generally inaccessible truncation error $d_{\rm MB}$.
As correlations strengthen, subleading contributions can be included systematically using the series expansion above. 
The residual spread can subsequently be used to calibrate predictive uncertainties.

\begin{figure}[t]
    \centering
    \includegraphics[width=0.8\linewidth]{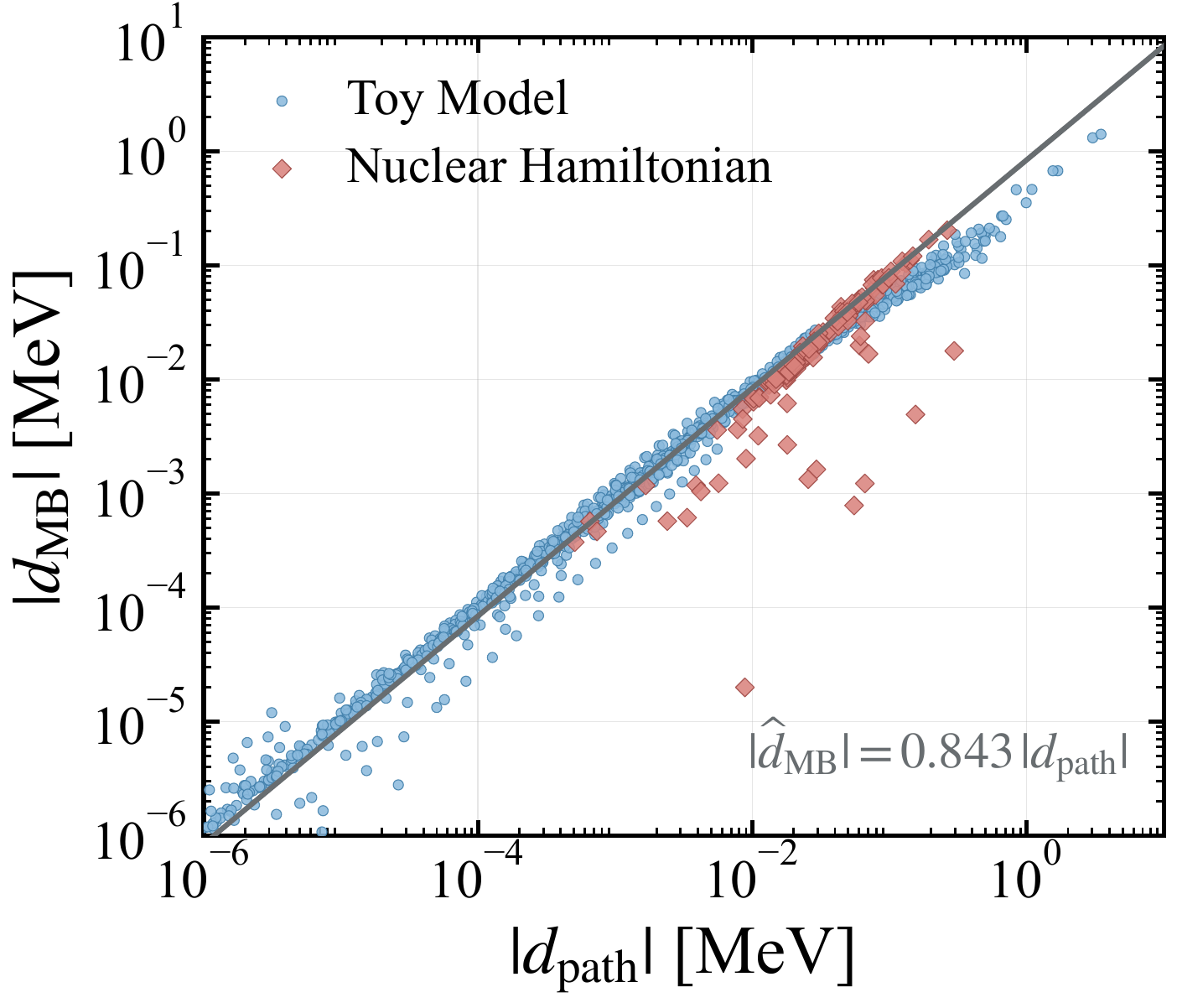}
    \caption{\label{fig:fourth}
Leading-order linear relation between the forward--backward path discrepancy and the IMSRG(2) many-body error for 1178 toy-model calculations and 126 calculations with nuclear Hamiltonians. Both sets span diverse Hamiltonians and model spaces. The gray line shows the proportional fit.}
\end{figure}

Fig.~\ref{fig:fourth} tests this relation using 1178 calculations from 16 toy-model Hamiltonian families and 126 calculations from six families of nuclear interactions. The toy models cover four different setups of model spaces and representative Hamiltonians from both condensed matter and nuclear physics. 
The nuclear calculations span several momentum spaces and use direct diagonalization or FCIQMC calculations to obtain the benchmark correlation energies. 
All replays use the TS closure. 
The errors follow a common proportional trend over more than six orders of magnitude. 
A robust fit to the logarithmic ratio gives
\begin{align}
|\widehat d_{\rm MB}|=0.843\,|d_{\rm path}|. 
\end{align}
The correlation across model systems and nuclear Hamiltonians is consistent with errors beginning at the same order in the omitted many-body contributions.
The remaining spread reflects Hamiltonian-dependent coefficients and higher-order effects. 
Calculation details and further diagnostics are given in the Supplemental Material~\cite{SM}. 
A tentative Bayesian model presented in the \hyperref[sec:end_matter]{End Matter} yields promising error estimates in good agreement with quasi-exact FCIQMC calculations in large space.
These benchmarks support replay as a useful probe of many-body uncertainties and motivate further study.

\textit{Conclusion.---}
We have introduced IMSRG(2)-BBGKY(2), which reconstructs complete one- and two-body RDMs while preserving the $\mathcal{O}(N_{\rm sp}^{6})$ scaling of IMSRG(2). 
In neutron matter, the reconstructed RDMs reveal that Fermi-sea depletion can vary non-monotonically with density and trace the contrasting depletion trends of the two interactions primarily to spin-triplet correlations.
A single collective replay provides access to the correlation structure encoded in RDMs, while the forward--backward discrepancy supplies an internal diagnostic of IMSRG(2) truncation errors. 
This framework is especially valuable for strongly correlated nuclear systems, in which energy alone does not fully characterize the underlying correlations. 
The reconstructed RDMs could also provide correlated initial conditions for time-dependent density-matrix calculations, enabling studies of how initial correlations influence subsequent dynamics.
Extensions to IMSRG(2$^\ast$) and IMSRG(3f2) and further applications to symmetric nuclear matter and finite nuclei are natural next steps~\cite{Stroberg2024PRC, He2024PRC}.

\textit{Acknowledgments.---} 
The authors would like to thank R.~Z.~Hu for discussions and providing the FCIQMC results of nuclear matter.
Thanks to S.~C.~Li, J.~W.~Chen and X.~Wang for useful discussions. 
This work is supported by the National Key R\&D Program of China under Grant Nos.~2023YFA1606400 and 2024YFA1610900; the National Natural Science Foundation of China under Grant Nos.~12335007,~12535008 and 12475118. 
We acknowledge the High-Performance Computing Platform of Peking University for providing computational resources.

\textit{Data Availability.} The processed data required to reproduce all figures in the main text are publicly available in the GitHub repository~\cite{data_git}. The complete raw datasets are available from the corresponding author upon reasonable request.

\bibliographystyle{modified-apsrev4-2.bst}
\bibliography{reference}

\clearpage

\section*{End Matter}
\label{sec:end_matter}
To illustrate how IMSRG(2)-BBGKY(2) replay can quantify many-body uncertainties in practical large-space calculations, we calibrate the path diagnostic using the two interactions considered in the main text. For each interaction, we model the dimensionless ratio $R_E=|d_{\rm MB}|/|d_{\rm path}|$, using standard statistical methods. All energies are evaluated for the same Hamiltonian and model space.
The calibration uses three model spaces, $(A,N_{\rm sp})=(14,38)$, $(14,54)$, and $(38,54)$, as included in Fig.~\ref{fig:fourth} of the main text.
Each space contains four matched calculations at $\rho=0.08$, $0.12$, $0.16$, and $0.20~\mathrm{fm}^{-3}$, yielding twelve calibration points per interaction.
The FCIQMC statistical uncertainties are negligible in these calculations with more than $10^7$ walkers and are therefore neglected.
The larger space, $(A,N_{\rm sp})=(38,514)$, is excluded from the calibration and serves as the prediction target.

For interaction $h$, density $i$, and model space $s$, we adopt~\cite{Gelman2013,Lange1989JASA}
\begin{align}
\log R_{E,his}
&=
\alpha_h+\beta_h x_i+u_{h,s}+\sigma_h t_{his},
\nonumber\\
u_{h,s}&\sim\mathcal{N}(0,\tau_h^2),
\qquad
t_{his}\sim t_{\nu=3}(0,1),
\nonumber\\
x_i&=\frac{\rho_i-\bar\rho}{s_\rho}.
\end{align}
The calibration densities give $\bar\rho=0.14~\mathrm{fm}^{-3}$ and $s_\rho=0.04472~\mathrm{fm}^{-3}$.
The coefficients $\alpha_h$ and $\beta_h$ describe the logarithmic center and its linear density dependence.
The random effect $u_{h,s}$ allows a common offset for calculations within the same model space, while $\sigma_h$ controls the residual Student-$t$ scale.
The two interactions are fitted independently, with priors~\cite{Gelman2013}
\begin{align}
\alpha_h,\beta_h\sim\mathcal{N}(0,2^2); \ \tau_h^2,\sigma_h^2\sim\operatorname{InvGamma}(2,0.20),
\end{align}
using the inverse-gamma convention $p(q|a,b)\propto q^{-(a+1)}\exp(-b/q)$.
For each interaction, we sample four chains of $30\,000$ iterations.
After discarding the first $8000$ iterations of each chain and retaining every fifth subsequent draw, we obtain $17\,600$ posterior samples, with rank-normalized split $\widehat R<1.001$~\cite{Vehtari2021}.
The posterior parameters are summarized in Table~\ref{tab:ed_parameters}.

\begin{table}[htbp]
\caption{\label{tab:ed_parameters}
Posterior medians and central $90\%$ credible intervals for the interaction-specific calibrations.}
\begin{ruledtabular}
\begin{tabular}{lcc}
Parameter & Texas & EMN\\
\hline
$\alpha_h$ &
$1.047\ [0.638,1.451]$ &
$0.587\ [0.047,1.123]$\\
$\beta_h$ &
$0.230\ [0.097,0.359]$ &
$-0.086\ [-0.244,0.054]$\\
$\sigma_h$ &
$0.223\ [0.161,0.332]$ &
$0.242\ [0.174,0.366]$\\
$\tau_h$ &
$0.366\ [0.230,0.674]$ &
$0.493\ [0.303,0.906]$\\
\end{tabular}
\end{ruledtabular}
\end{table}

For an unseen model space, we draw $u_{h,*}\sim\mathcal{N}(0,\tau_h^2)$ and $t_{h,*}\sim t_3(0,1)$ for each posterior sample and obtain
\begin{equation}
R_{E,h,*}(\rho)
=
\exp\!\left[
\alpha_h+\beta_h x(\rho)+u_{h,*}+\sigma_h t_{h,*}
\right].
\end{equation}
Thus, uncertainty associated with transfer to a new space is marginalized rather than setting its space-dependent shift to zero~\cite{Gelman2013}.
Let $q_{p,h}(\rho)$ denote the $p$-quantile of the predictive error magnitude $|d_{\rm path}|_{h,*}(\rho)R_{E,h,*}(\rho)$.
Since the model does not determine the error sign, we report the energy interval
\begin{equation}
\frac{E_{\rm IMSRG(2)}(\rho)}{A}
\pm q_{p,h}(\rho),
\qquad p=0.68,\ 0.90.
\end{equation}
These are pointwise predictive intervals incorporating parameter uncertainty, residual scatter, and variation among model spaces. 
We adopt this as a tentative error model, although other reasonable choices are possible.

\begin{figure}[htbp]
\centering
\includegraphics[width=0.9\linewidth]{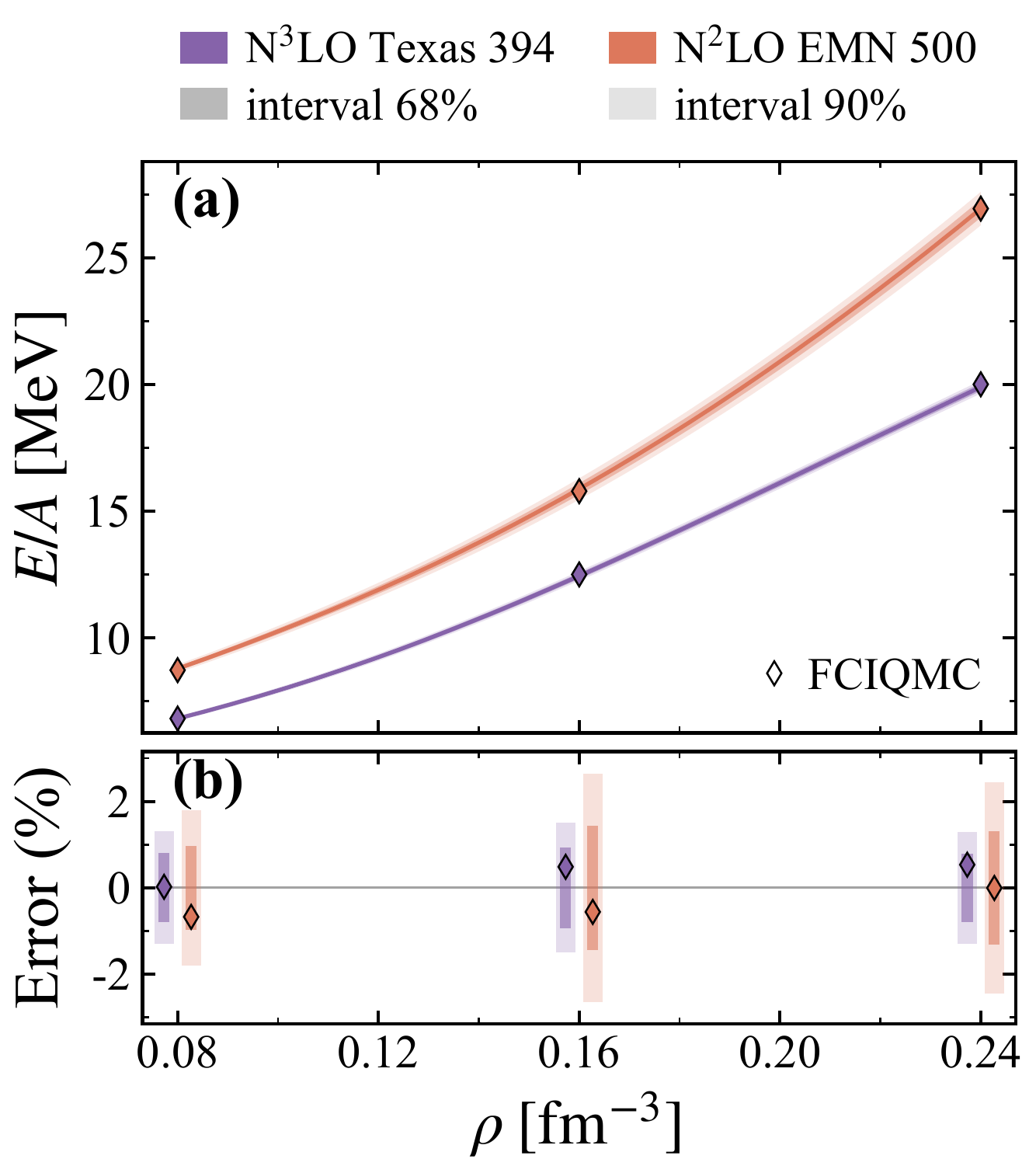}
\caption{\label{fig:ed_prediction}
Bayesian uncertainty estimates transferred to the larger neutron-matter space. (a) IMSRG(2) energies per particle with $68\%$ and $90\%$ predictive intervals. Diamonds denote held-out FCIQMC results. (b) Percentage differences between FCIQMC and IMSRG(2), with the corresponding predictive intervals centered at zero.}
\end{figure}

Fig.~\ref{fig:ed_prediction} shows the predictions obtained from the calculated path discrepancies in the larger model space.
All six held-out FCIQMC energies lie within the corresponding $68\%$ intervals.
The prediction at $\rho=0.24~\mathrm{fm}^{-3}$ also extends beyond the calibration density range.
The intervals estimate the many-body error for the specified NO2B Hamiltonian and do not include uncertainties from the nuclear interaction, the NO2B approximation, or finite-volume effects.
This analysis illustrates the potential of IMSRG(2)-BBGKY(2) replay to assess otherwise unknown IMSRG truncation uncertainties.

\end{document}


\title{Supplemental Material for: Reduced Density Matrices from Renormalization-Group Flows}

\author{X. Zhen\,\orcidlink{0009-0000-1806-4123}}
\affiliation{School of Physics, and State Key Laboratory of Nuclear Physics and Technology, Peking University, Beijing 100871, China}
\author{Z.~Y.~Meng\,\orcidlink{0009-0007-2538-7706}}
\affiliation{School of Physics, and State Key Laboratory of Nuclear Physics and Technology, Peking University, Beijing 100871, China}
\author{J. C. Pei\,\orcidlink{0000-0002-9286-1304}}\email[]{peij@pku.edu.cn}
\affiliation{School of Physics, and State Key Laboratory of Nuclear Physics and Technology, Peking University, Beijing 100871, China}
\affiliation{Southern Center for Nuclear-Science Theory (SCNT), Institute of Modern Physics, Chinese Academy of Sciences, Huizhou 516000, China}
\author{F. R. Xu\,\orcidlink{0000-0001-6699-0965}}\email[]{frxu@pku.edu.cn}
\affiliation{School of Physics, and State Key Laboratory of Nuclear Physics and Technology, Peking University, Beijing 100871, China}
\affiliation{Southern Center for Nuclear-Science Theory (SCNT), Institute of Modern Physics, Chinese Academy of Sciences, Huizhou 516000, China}

\maketitle

\section{Details of the theory}
In this section, we provide necessary theoretical details omitted in the main text. 
We only restate the definitions needed to fix the conventions used. 
Unless stated otherwise, all many-body operator matrix elements are antisymmetrized with respect to permutations of the bra and ket indices separately.

\subsection{The IMSRG(2)--BBGKY(2) flow equations}

The IMSRG flow is defined by~\cite{Tsukiyama2011PRL,Hergert2016PhysRep}
\begin{align}
    H(s) = U(s) H(0) U^\dagger(s), \
    \frac{{\rm d}H(s)}{{\rm d}s} = [\eta(s),H(s)],
    \label{eq:flow}
\end{align}
where
\begin{equation}
    \eta(s)
    \equiv
    \frac{{\rm d}U(s)}{{\rm d}s}U^\dagger(s),
    \qquad
    \eta^\dagger(s)=-\eta(s).
\end{equation}
The commutator of two particle-number-conserving normal-ordered operators of ranks $m$ and $n$ generally contains components up to rank $m+n-1$,
\begin{equation}
    [A^{(m)},B^{(n)}]
    =
    \sum_{k=|m-n|}^{m+n-1} C^{(k)}.
\end{equation}
At the IMSRG(2) level, both the Hamiltonian and the generator are truncated at the normal-ordered two-body (NO2B) level throughout the flow,
\begin{equation}
    H(s)=E(s)+f(s)+\Gamma(s),
    \qquad
    \eta(s)=\eta^{(1)}(s)+\eta^{(2)}(s).
\end{equation}
All intrinsic and induced normal-ordered operators of rank three and higher are discarded. 
For instance, projecting Eq.~(\ref{eq:flow}) onto its two-body component gives
\begin{align}
    \frac{{\rm d}\Gamma}{{\rm d}s}
    ={}&
    [\eta^{(1)},\Gamma]^{(2)}
    +[\eta^{(2)},f]^{(2)}
    +[\eta^{(2)},\Gamma]^{(2)} .
\end{align}
Here, the superscript denotes the NO2B component of the corresponding commutator. 
Applying Wick's theorem and collecting terms of equal particle rank yields the usual component form of the IMSRG(2) flow equations~\cite{Shavitt2009CUP,Hergert2016PhysRep}. 
We introduce this structure only to establish the notation and to facilitate comparison with the BBGKY(2) flow equations below. 
Note that these well-known formulas are valid only when all matrix elements are real, which is not the case for nuclear matter in plane-wave basis. For our implementation, we instead calculate all necessary commutators directly.\\
\textbf{IMSRG(2) flow equations:}
\begin{align}
    \frac{{\rm d}E}{{\rm d}s}
    ={}&
    \sum_{ab}
    (n_a-n_b)\,
    \eta_{ab}f_{ba}+
    \frac{1}{2}
    \sum_{abcd}
    n_a n_b \bar n_c \bar n_d\,
    \eta_{abcd}\Gamma_{cdab},
    \\[1ex]
    \frac{{\rm d}f_{12}}{{\rm d}s}
    ={}&
    \sum_a
    \left(
        \eta_{1a}f_{a2}
        -
        f_{1a}\eta_{a2}
    \right)+
    \sum_{ab}
    (n_a-n_b)
    \left(
        \eta_{ab}\Gamma_{b1a2}
        -
        f_{ab}\eta_{b1a2}
    \right)
    \notag\\
    &+
    \frac{1}{2}
    \sum_{abc}
    \left(
        n_a n_b \bar n_c
        +
        \bar n_a \bar n_b n_c
    \right)
    \left(
        \eta_{c1ab}\Gamma_{abc2}
        -
        \Gamma_{c1ab}\eta_{abc2}
    \right),
    \\[1ex]
    \frac{{\rm d}\Gamma_{1234}}{{\rm d}s}
    ={}&
    \sum_a
    (1-P_{12})
    \left(
        \eta_{1a}\Gamma_{a234}
        -
        f_{1a}\eta_{a234}
    \right)-
    \sum_a
    (1-P_{34})
    \left(
        \eta_{a3}\Gamma_{12a4}
        -
        f_{a3}\eta_{12a4}
    \right)
    \notag\\
    &+
    \frac{1}{2}
    \sum_{ab}
    (1-n_a-n_b)
    \left(
        \eta_{12ab}\Gamma_{ab34}
        -
        \Gamma_{12ab}\eta_{ab34}
    \right)-
    \sum_{ab}
    (n_a-n_b)
    (1-P_{12})(1-P_{34})\,
    \eta_{b2a4}\Gamma_{a1b3}.
\end{align}
Here, $n_a$ denotes the occupation number of the reference state, $\bar n_a\equiv 1-n_a$, and $P_{ij}$ exchanges the external indices $i$ and $j$. 
All explicit $s$ dependencies have been suppressed for brevity.

On the other hand, the RDM equations of motion are formulated using operators normal-ordered with respect to the physical vacuum. Since the IMSRG generator is stored in a form normal-ordered with respect to the reference state, we first transform it to the vacuum-normal-ordered representation according to Eq.~(4) of the main text. 
For notational simplicity, we omit the tilde of $\widetilde{\eta}$ in the BBGKY(2) equations below. The resulting generator is written as
\begin{equation}
    \eta
    =
    \sum_{ab}
    \eta^{(1)}_{ab}a_a^\dagger a_b
    +
    \frac{1}{4}
    \sum_{abcd}
    \eta^{(2)}_{abcd}
    a_a^\dagger a_b^\dagger a_d a_c .
\end{equation}
The back-transformed generator remains antisymmetric and anti-Hermitian. 
For backward evolution, we introduce the reversed flow parameter $\tau$ and define
\begin{equation}
    \eta_{\mathrm{rev}}(\tau)
    =
    -\eta(s_f-\tau), \tau \in [0, s_f].
    \label{eq:reverse_generator}
\end{equation}
In the following, $\eta$ denotes $\eta_{\mathrm{rev}}(\tau)$, and its explicit $\tau$ dependence is suppressed. 
Since the reversal of the flow is already included in Eq.~(\ref{eq:reverse_generator}), no additional minus sign is introduced in the RDM equations below.
Let $\varrho^{(N)}$ denote the full $N$-body density operator. With the commutator convention used in our implementation, its backward evolution satisfies
\begin{equation}
    \dot{\varrho}^{(N)}
    =
    [\eta,\varrho^{(N)}].
    \label{eq:density_flow}
\end{equation}
The $k$-body RDM is defined as~\cite{Coleman1963RMP,Lackner2015PRA}
\begin{equation}
    \varrho^{(k)}_{1\cdots k}
    =
    \frac{N!}{(N-k)!}
    \operatorname{Tr}_{k+1,\ldots,N}\varrho^{(N)},
\end{equation}
which is normalized to the particle number. With this convention we have
\begin{equation}
    \operatorname{Tr}_2\varrho^{(2)}_{12}
    =
    (N-1)\varrho^{(1)}_1,
    \qquad
    \operatorname{Tr}_3\varrho^{(3)}_{123}
    =
    (N-2)\varrho^{(2)}_{12},\qquad \text{etc.}
    \label{eq:contraction}
\end{equation}
In first-quantized notation, the generator acting in the full $N$-particle space is
\begin{equation}
    \eta
    =
    \sum_{i=1}^{N}\eta^{(1)}_i
    +
    \sum_{i<j}^{N}\eta^{(2)}_{ij}.
\end{equation}
Taking the corresponding partial traces of Eq.~(\ref{eq:density_flow}) gives the first two equations of the BBGKY hierarchy~\cite{Born1946PRSA,Bogoliubov1946JPhysUSSR,Kirkwood1946JCP,Yvon1935Hermann,Akbari2012PRB}:
\begin{equation}
    \dot{\varrho}^{(1)}_1
    =
    [\eta^{(1)}_1,\varrho^{(1)}_1]
    +
    \operatorname{Tr}_2
    [\eta^{(2)}_{12},\varrho^{(2)}_{12}]
    \label{eq:flow_1dm}
\end{equation}
and
\begin{equation}
    \begin{aligned}
        \dot{\varrho}^{(2)}_{12}
        ={}&
        \left[
        \eta^{(1)}_1
        +\eta^{(1)}_2
        +\eta^{(2)}_{12},
        \varrho^{(2)}_{12}
        \right]+
        \operatorname{Tr}_3
        \left[
        \eta^{(2)}_{13}
        +\eta^{(2)}_{23},
        \varrho^{(3)}_{123}
        \right].
    \end{aligned}
    \label{eq:flow_2dm}
\end{equation}
The subscripts $1$, $2$, and $3$ identify particle slots rather than single-particle indices.
Eq.~(\ref{eq:flow_1dm}) is closed at the two-body level. In Eq.~(\ref{eq:flow_2dm}), only the final term, referred to as the spectator contribution, depends on the 3-RDM. 
Similarly, the operators $\eta^{(1)}_1$ and $\eta^{(1)}_2$ are different embeddings of the same vacuum-normal-ordered one-body generator.
Before restriction to the antisymmetric subspace, their matrix elements in a tensor-product basis are
\begin{align}
    \langle pq|\eta^{(1)}_1|rs\rangle
    =
    \eta^{(1)}_{pr}\delta_{qs},\ \langle pq|\eta^{(1)}_2|rs\rangle
    =
    \delta_{pr}\eta^{(1)}_{qs}.
\end{align}
Likewise, $\eta^{(2)}_{12}$, $\eta^{(2)}_{13}$, and $\eta^{(2)}_{23}$ are different embeddings of the same two-body tensor $\eta^{(2)}_{abcd}$. In an unsymmetrized three-particle tensor-product basis,
\begin{align}
    \langle pqr|\eta^{(2)}_{12}|stu\rangle
    =
    \eta^{(2)}_{pqst}\delta_{ru},\ 
    \langle pqr|\eta^{(2)}_{13}|stu\rangle
    =
    \eta^{(2)}_{prsu}\delta_{qt},\ 
    \langle pqr|\eta^{(2)}_{23}|stu\rangle
    =
    \delta_{ps}\eta^{(2)}_{qrtu}.
\end{align}
In the second-quantized component equations, all exchange signs
associated with this restriction are generated automatically by the
fermionic anticommutation relations. 
Following the conventions of the main text, the RDM components used in our implementation are defined as
\begin{align}
    \varrho^{(1)}_{pq}
    =
    \langle a_p^\dagger a_q\rangle,
    \varrho^{(2)}_{pqrs}
    =
    \langle
    a_p^\dagger a_q^\dagger a_s a_r
    \rangle,
    \varrho^{(3)}_{pqrstu}
    =
    \langle
    a_p^\dagger a_q^\dagger a_r^\dagger
    a_u a_t a_s
    \rangle.
    \label{eq:rdm_comp}
\end{align}
For a $k$-body RDM, the first $k$ indices label the creation operators, whereas the last $k$ indices label the annihilation operators in reverse order. 
To illustrate how the index combinations in the component equations arise, consider the part of
\begin{equation}
    \left\langle
    \left[
    a_p^\dagger a_q^\dagger a_s a_r,
    \eta^{(2)}
    \right]
    \right\rangle
\end{equation}
in which $a_r$ is contracted with a creation operator in the two-body generator. Before performing the contractions, the relevant part of the product in the first ordering of the commutator is
\begin{equation}
    \frac{1}{4}
    \sum_{\mu\nu bc}
    \eta^{(2)}_{\mu\nu bc}
    a_p^\dagger a_q^\dagger a_s
    a_r
    a_\mu^\dagger a_\nu^\dagger
    a_c a_b .
    \label{eq:example_product}
\end{equation}
Using the canonical anticommutation relation
\begin{equation}
    \{a_i,a_j^\dagger\}
    =
    \delta_{ij},
\end{equation}
we rewrite Eq.~(\ref{eq:example_product}) as
\begin{align}
    \frac{1}{4}
    \sum_{\mu\nu bc}
    \eta^{(2)}_{\mu\nu,bc}
    a_p^\dagger a_q^\dagger a_s
    \Big(
        \delta_{r\mu}a_\nu^\dagger
        -
        \delta_{r\nu}a_\mu^\dagger
        +
        a_\mu^\dagger a_\nu^\dagger a_r
    \Big)
    a_c a_b .
\end{align}
Retaining the single-contraction terms, taking their expectation values, and using Eq.~(\ref{eq:rdm_comp}), we obtain
\begin{align}
    -\frac{1}{4}
    \sum_{\mu\nu bc}
    \eta^{(2)}_{\mu\nu bc}
    \left(
        \delta_{r\mu}
        \varrho^{(3)}_{pq\nu bcs}
        -
        \delta_{r\nu}
        \varrho^{(3)}_{pq\mu bcs}
    \right).
\end{align}
Using the antisymmetric properties,
\begin{align}
    -\frac{1}{4}
    \sum_{\mu\nu bc}
    \eta^{(2)}_{\mu\nu bc}
    \left(
        \delta_{r\mu}
        \varrho^{(3)}_{pq\nu bcs}
        -
        \delta_{r\nu}
        \varrho^{(3)}_{pq\mu bcs}
    \right)=
    -\frac{1}{4}
    \sum_{\nu bc}
    \eta^{(2)}_{r\nu bc}
    \varrho^{(3)}_{pq\nu bcs}
    +
    \frac{1}{4}
    \sum_{\mu bc}
    \eta^{(2)}_{\mu r bc}
    \varrho^{(3)}_{pq\mu bcs}=
    -\frac{1}{2}
    \sum_{abc}
    \eta^{(2)}_{ra bc}
    \varrho^{(3)}_{pqa bcs}.
\end{align}
This is the first spectator term in the complete 2-RDM equation.
The other three spectator terms follow from the corresponding single contractions involving $a_s$, $a_p^\dagger$, and $a_q^\dagger$. The zero-contraction contributions cancel between the two operator orderings in the commutator, while the double contractions generate the internal two-particle contribution. Repeating this procedure for all possible contractions yields the component form of the BBGKY(2) flow equations, whose structure closely parallels that of the IMSRG(2) equations.\\
\textbf{BBGKY(2) flow equations:}
\begin{align}
    \dot{\varrho}^{(1)}_{pq}
        ={}&
        \sum_a
        \left(
            \eta^{(1)}_{qa}\varrho^{(1)}_{pa}
            -
            \eta^{(1)}_{ap}\varrho^{(1)}_{aq}
        \right)+
        \frac{1}{2}
        \sum_{abc}
        \left(
            \eta^{(2)}_{qabc}\varrho^{(2)}_{pabc}
            -
            \eta^{(2)}_{abpc}\varrho^{(2)}_{abqc}
        \right),\\
    \dot{\varrho}^{(2)}_{pqrs}
        ={}&
        \sum_a\Big[
        -\eta^{(1)}_{ap}\varrho^{(2)}_{aqrs}
        +\eta^{(1)}_{aq}\varrho^{(2)}_{aprs}
        +\eta^{(1)}_{sa}\varrho^{(2)}_{pqra}
        +\eta^{(1)}_{ra}\varrho^{(2)}_{pqas}
        \Big]\nonumber
        \\
        &+
        \frac{1}{2}
        \sum_{ab}
        \left(
            \eta^{(2)}_{rsab}\varrho^{(2)}_{pqab}
            -
            \eta^{(2)}_{abpq}\varrho^{(2)}_{abrs}
        \right)\nonumber
        \\
        &+
        \frac{1}{2}
        \sum_{abc}\Big[
        -\eta^{(2)}_{rabc}
         \varrho^{(3)}_{pqabcs}
        +
        \eta^{(2)}_{sabc}
         \varrho^{(3)}_{pqabcr} +
        \eta^{(2)}_{abpc}
         \varrho^{(3)}_{abqrsc}
        -
        \eta^{(2)}_{abqc}
         \varrho^{(3)}_{abprsc}
        \Big].        
\end{align}
Compared with the IMSRG(2) equations, the BBGKY(2) equations contain no explicit reference-state occupation factors, and the corresponding particle-hole ring structure does not appear. 
These contractions hence can be efficiently optimized by matrix multiplication.  
At this stage, however, the hierarchy remains open because the 2-RDM equation depends explicitly on $\varrho^{(3)}$, analogous to the appearance of induced three-body operators in the IMSRG flow. A closed BBGKY(2) approximation is obtained by reconstructing $\varrho^{(3)}$ exclusively from $\varrho^{(1)}$ and $\varrho^{(2)}$. The reconstruction strategies employed in this work are discussed in the next subsection.

\subsection{Closure strategies}

In general, a higher-order RDM can be decomposed into antisymmetrized products of lower-order RDMs and an irreducible connected contribution, referred to as the cumulant. Closing the BBGKY hierarchy at the two-body level therefore requires reconstructing the 3-RDM from the instantaneous 1-RDM and 2-RDM, together with an approximation to the genuine three-body cumulant. We employ the standard Grassmann wedge product~\cite{Harriman2002PRA,Akbari2012PRB,Lackner2015PRA}. Let
\begin{equation}
\widehat{\mathcal A}_k
=
\frac{1}{k!}
\sum_{\pi\in S_k}
\operatorname{sgn}(\pi)\widehat P_\pi
\end{equation}
denote the normalized projector onto the antisymmetric $k$-particle space, where $\widehat P_\pi$ permutes the particle slots according to $\pi$. For a $p$-body operator $A^{(p)}$ and a $q$-body operator $B^{(q)}$, their Grassmann product is defined by
\begin{equation}
A^{(p)}\wedge B^{(q)}
=
\widehat{\mathcal A}_{p+q}
\left(
A^{(p)}\otimes B^{(q)}
\right)
\widehat{\mathcal A}_{p+q}.
\label{eq:grassmann_product}
\end{equation}
In a single-particle basis, this becomes
\begin{align}
&
\left(
A^{(p)}\wedge B^{(q)}
\right)_
{p_1\cdots p_{p+q},q_1\cdots q_{p+q}}=
\frac{1}{[(p+q)!]^2}
\sum_{\pi,\sigma\in S_{p+q}}
\operatorname{sgn}(\pi)
\operatorname{sgn}(\sigma)
A_{
p_{\pi(1)}\cdots p_{\pi(p)},
q_{\sigma(1)}\cdots q_{\sigma(p)}
}
B_{
p_{\pi(p+1)}\cdots p_{\pi(p+q)},
q_{\sigma(p+1)}\cdots q_{\sigma(p+q)}
}.
\end{align}
This product is bilinear, commutative, and associative. 
The numerical factors appearing below result from the normalization of the antisymmetrizers in Eq.~\eqref{eq:grassmann_product}.
For two one-body matrices,
\begin{equation}
2
\left(
\varrho^{(1)}\wedge\varrho^{(1)}
\right)_{pqrs}
=
\varrho^{(1)}_{pr}\varrho^{(1)}_{qs}
-
\varrho^{(1)}_{ps}\varrho^{(1)}_{qr}.
\end{equation}
The two-body density cumulant is consequently defined as
\begin{equation}
\lambda^{(2)}_{12}
=
\varrho^{(2)}_{12}
-
2\varrho^{(1)}_1\wedge\varrho^{(1)}_2,
\label{eq:lambda2}
\end{equation}
or, in the single-particle basis,
\begin{equation}
\lambda^{(2)}_{pqrs}
=
\varrho^{(2)}_{pqrs}
-
\varrho^{(1)}_{pr}\varrho^{(1)}_{qs}
+
\varrho^{(1)}_{ps}\varrho^{(1)}_{qr}.
\end{equation}
With the same convention, the cumulant expansion of the 3-RDM is
\begin{align}
\varrho^{(3)}_{123}
={}&
6\,
\varrho^{(1)}_1
\wedge
\varrho^{(1)}_2
\wedge
\varrho^{(1)}_3
+
9\,
\lambda^{(2)}_{12}
\wedge
\varrho^{(1)}_3
+
\lambda^{(3)}_{123}.
\end{align}
The first term is the independent-particle contribution, the second describes two-body correlations in the presence of a spectator particle, and $\lambda^{(3)}$ contains the genuine connected three-body correlations. The closure strategies considered below differ only in their approximation to $\lambda^{(3)}$.

The Valdemoro (V) reconstruction neglects the genuine three-body cumulant~\cite{Colmenero1993PRA},
\begin{equation}
\lambda^{(3)}_{\rm V}=0.
\end{equation}
The corresponding reconstructed 3-RDM is
\begin{align}
\varrho^{(3)}_{\rm V}
={}&
6\,
\varrho^{(1)}_1
\wedge
\varrho^{(1)}_2
\wedge
\varrho^{(1)}_3
+
9\,
\lambda^{(2)}_{12}
\wedge
\varrho^{(1)}_3.
\end{align}
Equivalently, after eliminating $\lambda^{(2)}$ using Eq.~\eqref{eq:lambda2},
\begin{align}
\varrho^{(3)}_{\rm V}
={}&
9\,
\varrho^{(2)}_{12}
\wedge
\varrho^{(1)}_3
-
12\,
\varrho^{(1)}_1
\wedge
\varrho^{(1)}_2
\wedge
\varrho^{(1)}_3.
\end{align}
The Nakatsuji--Yasuda (NY) reconstruction includes an approximation to the three-body cumulant that is quadratic in $\lambda^{(2)}$~\cite{Nakatsuji1996PRL,Yasuda1997PRA}. It contains an internal contraction between two two-body cumulants and is therefore not an ordinary Grassmann product of independent particle-space operators. We first define the unantisymmetrized contracted tensor
\begin{equation}
K_{123}
=
\lambda^{(2)}_{12}
\mathcal P_2
\lambda^{(2)}_{23},
\end{equation}
whose matrix elements are
\begin{equation}
\left(K_{123}\right)_{pqtrsu}
=
\sum_{ab}
\lambda^{(2)}_{pqra}
\mathcal P_{ab}
\lambda^{(2)}_{btsu}.
\label{eq:ny_seed}
\end{equation}

The NY three-body cumulant is the fully antisymmetrized contracted product
\begin{equation}
\lambda^{(3)}_{\rm NY}
=
9\,
\widehat{\mathcal A}_3
K_{123}
\widehat{\mathcal A}_3
.
\label{eq:lambda3_ny}
\end{equation}
The factor $9$ compensates the normalization of the left and right three-particle antisymmetrizers. It gives each of the nine inequivalent placements of the spectator bra and ket indices unit weight. More explicitly, let
\begin{equation}
P=(p_1,p_2,p_3),
\qquad
Q=(q_1,q_2,q_3),
\end{equation}
and let $P_{\widehat i}$ and $Q_{\widehat j}$ denote the ordered pairs obtained by deleting $p_i$ and $q_j$, respectively. Equation~\eqref{eq:lambda3_ny} is then
\begin{align}
\left(
\lambda^{(3)}_{\rm NY}
\right)_{P,Q}
={}&
\sum_{i=1}^{3}
\sum_{j=1}^{3}
(-1)^{i+j}
\sum_{ab}
\lambda^{(2)}_{P_{\widehat i},q_j a}
\mathcal P_{ab}
\lambda^{(2)}_{b p_i,Q_{\widehat j}}.
\label{eq:lambda3_ny_nine_terms}
\end{align}
This is the nine-term expression evaluated in the numerical implementation and does not require constructing a rank-six tensor. The one-body intermediate entering Eqs.~\eqref{eq:ny_seed}--\eqref{eq:lambda3_ny_nine_terms} is~\cite{Donsa2023PRResearch}
\begin{equation}
\mathcal P
=
\left(
3\Pi-I-\varrho^{(1)}
\right)^{-1}.
\end{equation}
Both $\Pi$ and $\mathcal P$ are constructed from the instantaneous 1-RDM at each point of the backward flow. We first diagonalize
\begin{equation}
\varrho^{(1)}
=
\sum_{\mu}
n_\mu
\ket{\phi_\mu}
\bra{\phi_\mu},
\qquad
n_1\geq n_2\geq\cdots,
\end{equation}
where $n_\mu$ and $\ket{\phi_\mu}$ are the natural occupations and natural orbitals. This is particularly feasible in nuclear matter with the conservation of momentum. The rank-$N$ occupation intermediate is
\begin{equation}
\Pi
=
\sum_{\mu}
\pi_\mu
\ket{\phi_\mu}
\bra{\phi_\mu},
\qquad
\pi_\mu
=
\begin{cases}
1, & \mu\leq N,\\
0, & \mu>N.
\end{cases}
\end{equation}
Thus, $\Pi$ projects onto the subspace spanned by the $N$ most occupied natural orbitals. Since $\Pi$ and $\varrho^{(1)}$ are diagonal in the same basis,
\begin{equation}
\mathcal P
=
\sum_{\mu}
\frac{1}
{3\pi_\mu-1-n_\mu}
\ket{\phi_\mu}
\bra{\phi_\mu}.
\end{equation}
For a physical 1-RDM with $0\leq n_\mu\leq1$, the denominators remain finite. If $n_N=n_{N+1}$, however, the rank-$N$ intermediate is not uniquely defined, and the NY reconstruction may become basis dependent within the degenerate natural-orbital subspace. 
In our calculations, the corresponding occupation gap remains finite and is monitored explicitly.
The complete NY reconstruction is
\begin{equation}
\varrho^{(3)}_{\rm NY}
=
\varrho^{(3)}_{\rm V}
+
\lambda^{(3)}_{\rm NY}.
\end{equation}

The Tohyama--Schuck (TS) reconstruction is motivated by a doubles-based approximation to the correlated many-body state~\cite{Tohyama2014EPJA}. This closure retains the connected two-particle--one-hole ($2p1h$) and one-particle--two-hole ($1p2h$) blocks that are factorized into products of two-body cumulants. The particle and hole orbitals are defined with respect to the original Slater-determinant reference $\ket{\Phi}$ and are kept fixed throughout the backward flow. For the $2p1h$ block, the TS reconstruction reads
\begin{equation}
\lambda^{(3)}_{
{\rm TS},p_1p_2h_1p_3p_4h_2
}
=
\frac{1}{\mathcal N_{\rm TS}}
\sum_h
\lambda^{(2)}_{hh_1p_3p_4}
\lambda^{(2)}_{p_1p_2h_2h}.
\end{equation}
Similarly, the $1p2h$ block is reconstructed as
\begin{equation}
\lambda^{(3)}_{
{\rm TS},p_1h_1h_2p_2h_3h_4
}
=
\frac{1}{\mathcal N_{\rm TS}}
\sum_p
\lambda^{(2)}_{h_1h_2p_2p}
\lambda^{(2)}_{p_1p h_3h_4}.
\end{equation}
Matrix elements with different index orderings follow from fermionic antisymmetry, while the remaining blocks of the connected three-body cumulant are set to zero. The normalization factor is~\cite{Tohyama2017EPJA}
\begin{align}
\mathcal N_{\rm TS}
={}&
1+
\sum_{h_1<h_2}
\sum_{p_1<p_2}
\left|
\lambda^{(2)}_{h_1h_2p_1p_2}
\right|^2=
1+
\frac{1}{4}
\sum_{h_1h_2p_1p_2}
\left|
\lambda^{(2)}_{h_1h_2p_1p_2}
\right|^2.
\end{align}
This factor approximately accounts for the normalization of the correlated doubles state and suppresses excessive growth of the reconstructed three-body cumulant when the $2p2h$ correlations become large. 
The complete TS reconstruction is
\begin{equation}
\varrho^{(3)}_{\rm TS}
=
\varrho^{(3)}_{\rm V}
+
\lambda^{(3)}_{\rm TS}.
\end{equation}

The three reconstruction strategies recover different contributions to the three-body correlations. The V reconstruction retains only the disconnected contributions determined by the 1-RDM and 2-RDM. The NY reconstruction supplements these terms by a fully antisymmetrized quadratic functional of the two-body cumulant. The TS reconstruction instead retains selected $2p1h$ and $1p2h$ sectors motivated by a doubles-based representation of the correlated state. Although both NY and TS are quadratic in $\lambda^{(2)}$, they therefore represent distinct approximations to the genuine three-body cumulant. 
Importantly, none of these closures requires the explicit construction or storage of the six-index 3-RDM. The reconstructed expressions are inserted directly into the spectator contraction appearing in the 2-RDM equation. As in the main text, we let $N_{\rm sp}$ denote the dimension of the single-particle space. A direct evaluation of
\begin{equation}
\sum_{abc}
\eta^{(2)}_{rabc}
\varrho^{(3)}_{pqabcs}
\end{equation}
from a stored 3-RDM would require $O(N_{\rm sp}^6)$ memory and $O(N_{\rm sp}^7)$ operations for all external indices. In the reconstructed expressions, the contraction can instead be ordered through lower-rank intermediates. For example, a representative V contribution can be evaluated as
\begin{equation}
\sum_{abc}
\eta^{(2)}_{rabc}
\lambda^{(2)}_{pqab}
\varrho^{(1)}_{cs}
=
\sum_{ab}
\lambda^{(2)}_{pqab}
\left[
\sum_c
\eta^{(2)}_{rabc}
\varrho^{(1)}_{cs}
\right].
\end{equation}
The bracketed four-index intermediate costs $O(N_{\rm sp}^5)$ to construct, while its contraction with $\lambda^{(2)}$ is a pair-space matrix multiplication with cost $O(N_{\rm sp}^6)$. A representative NY term has the structure
\begin{equation}
\sum_{abcxy}
\eta^{(2)}_{rabc}
\lambda^{(2)}_{qabx}
\mathcal P_{xy}
\lambda^{(2)}_{ypcs}.
\end{equation}
It can be evaluated in the order
\begin{align}
T_{rcqx}
&=
\sum_{ab}
\eta^{(2)}_{rabc}
\lambda^{(2)}_{qabx},
\\
\widetilde T_{rcqy}
&=
\sum_x
T_{rcqx}
\mathcal P_{xy},
\\
S_{pqrs}
&=
\sum_{cy}
\widetilde T_{rcqy}
\lambda^{(2)}_{ypcs}.
\end{align}
The first and final contractions scale as $O(N_{\rm sp}^6)$, the middle contraction scales as $O(N_{\rm sp}^5)$, and the largest intermediate contains only four single-particle indices. The other eight antisymmetrized NY contributions have the same maximal scaling. The factorized TS contributions admit analogous contraction orders and are no more expensive. Consequently, the V, NY, and TS closures all retain the leading
\begin{equation}
O(N_{\rm sp}^6)
\end{equation}
operation count of IMSRG(2), with $O(N_{\rm sp}^4)$ working memory~\cite{Hergert2016PhysRep}. The separately stored generator trajectory requires
$O(N_sN_{\rm sp}^4)$ storage for $N_s$ saved flow points, but it does not alter the polynomial scaling with the single-particle dimension. The closures avoid both the $O(N_{\rm sp}^6)$ storage of a complete 3-RDM and its explicit dynamical propagation. 
In general, the reconstructed 3-RDMs do not satisfy the exact contraction relation~\cite{Akbari2012PRB,Donsa2023PRResearch}, which motivates the contraction-consistent tangent projection introduced in the following subsection.

\subsection{Contraction-consistent tangent projection}
The particle-number-normalized RDMs should satisfy the exact contraction relation (\ref{eq:contraction})
\begin{equation}
\operatorname{Tr}_2
\varrho^{(2)}_{12}
=
(N-1)\varrho^{(1)}_1.
\notag
\end{equation}
Differentiating this equation with respect to the backward flow parameter gives the corresponding tangent-space condition
\begin{equation}
\operatorname{Tr}_2
\dot{\varrho}^{(2)}_{12}
=
(N-1)\dot{\varrho}^{(1)}_1.
\label{eq:tangent_contraction}
\end{equation}
Thus, an RDM trajectory that initially satisfies the contraction relation remains on the contraction-consistent manifold only if its instantaneous tangent vector satisfies Eq.~(\ref{eq:tangent_contraction}) throughout the evolution. For a chosen closure, let
\begin{equation}
\dot{\varrho}^{(1)}_{\rm raw},
\qquad
\dot{\varrho}^{(2)}_{\rm raw}
\end{equation}
denote the derivatives obtained directly from the BBGKY(2) equations before the projection. The 1-RDM equation is already closed in terms of $\varrho^{(1)}$ and $\varrho^{(2)}$, whereas the 2-RDM derivative depends on the reconstructed 3-RDM. We therefore leave the 1-RDM tangent unchanged and correct only the 2-RDM tangent. The instantaneous contraction defect is defined as
\begin{equation}
D
=
(N-1)\dot{\varrho}^{(1)}_{\rm raw}
-
\operatorname{Tr}_2
\dot{\varrho}^{(2)}_{\rm raw}.
\label{eq:tangent_defect}
\end{equation}
In the single-particle basis, this becomes
\begin{equation}
D_{pq}
=
(N-1)\dot{\varrho}^{(1)}_{{\rm raw},pq}
-
\sum_j
\dot{\varrho}^{(2)}_{{\rm raw},pjqj}.
\end{equation}
From the defect matrix, we construct a one-body tensor
\begin{equation}
\mathcal{X}
=
\frac{1}{N_{\rm sp}-2}
\left[
D
-
\frac{\operatorname{Tr}D}{2(N_{\rm sp}-1)}I
\right].
\label{eq:tangent_x}
\end{equation}
The correction to the 2-RDM tangent is then written using the wedge product as
\begin{equation}
\Delta\dot{\varrho}^{(2)}_{12}
=4\mathcal{X}_1\wedge I_2.
\end{equation}
Explicitly, its matrix elements are
\begin{align}
\Delta\dot{\varrho}^{(2)}_{pqrs}
={}&
\mathcal{X}_{pr}\delta_{qs}
-
\mathcal{X}_{ps}\delta_{qr}
-
\mathcal{X}_{qr}\delta_{ps}
+
\mathcal{X}_{qs}\delta_{pr}.
\label{eq:tangent_correction_component}
\end{align}
The projected tangent vector is therefore
\begin{align}
\dot{\varrho}^{(1)}
=
\dot{\varrho}^{(1)}_{\rm raw},
\dot{\varrho}^{(2)}
=
\dot{\varrho}^{(2)}_{\rm raw}
+
\Delta\dot{\varrho}^{(2)}.
\end{align}
To verify the contraction explicitly, contracting Eq.~(\ref{eq:tangent_correction_component}) over the second particle gives
\begin{align}
\sum_j
\Delta\dot{\varrho}^{(2)}_{pj qj}
&=
(N_{\rm sp}-2)\mathcal{X}_{pq}
+
\delta_{pq}\operatorname{Tr}\mathcal{X}.
\end{align}
Equation~(\ref{eq:tangent_x}) implies
\begin{equation}
\operatorname{Tr}\mathcal{X}
=
\frac{\operatorname{Tr}D}{2(N_{\rm sp}-1)},
\end{equation}
and consequently
\begin{equation}
\sum_j
\Delta\dot{\varrho}^{(2)}_{pj qj}
=
D_{pq}.
\label{eq:correction_equals_defect}
\end{equation}
Combining Eqs.~(\ref{eq:tangent_defect}) and
(\ref{eq:correction_equals_defect}) immediately yields
\begin{equation}
\operatorname{Tr}_2
\dot{\varrho}^{(2)}
=
(N-1)\dot{\varrho}^{(1)}.
\end{equation}
The correction in Eq.~(\ref{eq:tangent_correction_component}) is the minimum-norm antisymmetric two-body tensor whose contraction equals $D$~\cite{Harriman2002PRA,Lackner2015PRA}. Because the correction is constructed as a wedge product, it also preserves the fermionic antisymmetry of the 2-RDM tangent. In exact arithmetic, Hermiticity is likewise preserved whenever the raw BBGKY tangent is Hermitian. 

It is worth mentioning that this is not the only strategy to recover the contraction defects~\cite{Lackner2015PRA,Joost2022PRB}. The present projection corrects only the contraction-carrying component of the instantaneous 2-RDM tangent, while leaving its contraction-free component unchanged. 
More importantly, these closures may violate the necessary $N$-representability, which means that the resulting $\varrho$ may not correspond to a physical RDM~\cite{Coleman1963RMP,Akbari2012PRB}. In time-dependent density matrix theory, a purification procedure to retain a controlled $N$-representability is widely applied~\cite{Lackner2015PRA,Lackner2017PRA,Joost2022PRB}. 
We therefore monitor the magnitude of the tangent projection together with Hermiticity, particle-number traces, natural occupations, and other diagnostic variables, together with their dependence on the closure and IMSRG generators. These complementary diagnostics distinguish a numerically well-constrained reconstruction from one whose apparent consistency is maintained only by the projection. The details are illustrated in the next section.

\section{Systematic Diagnostics}

We employ a common set of diagnostics to assess the convergence of the Hamiltonian flow, the internal consistency of the reconstructed RDMs, and their sensitivity to the closure strategy and IMSRG generator. 
Unless stated otherwise, every diagnostic is evaluated independently for each closure and generator. Quantities involving a maximum over the flow parameter probe the complete reverse trajectory, whereas energy and RDM errors without an explicit flow argument refer to the final reconstructed state. The formulas below use the full ordered-index convention for the RDMs,
\begin{equation}
{\rm Tr}\varrho^{(1)}=A,
\qquad
{\rm Tr}\varrho^{(2)}=A(A-1),
\end{equation}
where $A$ denotes the number of nucleons. 
When a two-body tensor is stored in a canonical antisymmetric pair basis, its ordered-index trace is twice the pair-space trace and its ordered-index Frobenius norm is twice the pair-space Frobenius norm. 

\subsection{Scalable diagnostics}

The diagnostics introduced in this subsection require only the IMSRG(2) Hamiltonian trajectory and the reconstructed 1-RDM and 2-RDM. They are therefore available in calculations for which neither exact diagonalization nor fixed-$A$ Fock-space inversion is feasible.

For each generator, we denote the final IMSRG(2) energy by
\begin{equation}
E_{\rm IMSRG(2)}.
\end{equation}
For each closure $X\in\{{\rm V},{\rm NY},{\rm TS}\}$, the energy obtained from the reconstructed RDMs is~\cite{Lowdin1955PR,DePrince2024WCMS}
\begin{align}
E_{\rm BBGKY(2)}^{X}
={}
\sum_{pq}
h_{pq}^{(0)}
\varrho_{pq}^{(1),X}+
\frac{1}{4}
\sum_{pqrs}
V_{pqrs}^{(0)}
\varrho_{pqrs}^{(2),X}.
\end{align}
Here $h^{(0)}$ and $V^{(0)}$ define the initial bare Hamiltonian in physical-vacuum normal order. The Hamiltonian-side IMSRG evolution and the state-side RDM reconstruction provide two evaluations of the same energy. Their difference defines the path-consistency diagnostic
\begin{equation}
\Delta E_{{\rm path}}^{\rm X}
=
\frac{
\left|
E_{\rm BBGKY(2)}^{X}
-
E_{\rm IMSRG(2)}
\right|
}{A}.
\end{equation}
Unless a closure label is shown explicitly, we use the TS reconstruction,
\begin{equation}
\Delta E_{{\rm path}}
\equiv
\Delta E_{{\rm path}}^{\rm TS}.
\end{equation}
For an untruncated unitary transformation and an exact state reconstruction, the two energy evaluations coincide. At finite truncation, $\Delta E_{{\rm path}}$ measures the resulting Hamiltonian--density inconsistency. It provides a scalable internal estimate of the many-body-error scale after calibration against exactly solvable systems. The sensitivity of the reconstructed energy to the three-body closure is measured by
\begin{equation}
\Delta E_{\rm NY\text{-}TS}
=
\frac{
\left|
E_{\rm BBGKY(2)}^{\rm NY}
-
E_{\rm BBGKY(2)}^{\rm TS}
\right|
}{A},
\end{equation}
and by the full closure span
\begin{equation}
\Delta E_{\rm cl}
=
\frac{1}{A}
\left[
\max_X E_{\rm BBGKY(2)}^{X}
-
\min_X E_{\rm BBGKY(2)}^{X}
\right].
\end{equation}
A large closure span indicates that the available one- and two-body information does not determine the reconstructed three-body contribution robustly. When these energy differences are used as inputs to the uncertainty calibration, they enter through dimensionless ratios such as
\begin{equation}
\frac{\Delta E_{\rm NY\text{-}TS}}{\Delta E_{{\rm path}}},
\qquad
\frac{\Delta E_{\rm cl}}{\Delta E_{{\rm path}}}.
\end{equation}
Every closure directly provides the matrices
\begin{equation}
\varrho^{(1),X},
\qquad
\varrho^{(2),X}.
\end{equation}
Their closure dependence is quantified by the maximum symmetric pairwise Frobenius distance
\begin{equation}
\Delta R^{(k)}_{\rm cl}
=
\max_{X<X'}
\frac{
2
\left\|
\varrho^{(k),X}
-
\varrho^{(k),X'}
\right\|_{\rm F}
}{
\left\|
\varrho^{(k),X}
\right\|_{\rm F}
+
\left\|
\varrho^{(k),X'}
\right\|_{\rm F}
},
\qquad
k=1,2.
\end{equation}
Unlike an energy difference, $\Delta R^{(k)}_{\rm cl}$ probes the complete reconstructed matrix and is therefore less susceptible to cancellations among individual matrix elements. The Frobenius norm is
\begin{equation}
\left\|T^{(k)}\right\|_{\rm F}
=
\sqrt{
{\rm Tr}\left[
T^{(k)\dagger}T^{(k)}
\right]
}
=
\left[
\sum_{\alpha\beta}
\left|
T_{\alpha\beta}^{(k)}
\right|^2
\right]^{1/2}.
\end{equation}
The instantaneous strength of the two-body generator is measured by
\begin{equation}
\eta_2(s)
=
\left\|
\eta^{(2)}(s)
\right\|_{\rm F}.
\end{equation}
To compare calculations with different numbers of active two-body channels, we define
\begin{equation}
\overline{\eta}_2(s)
=
\frac{
\left\|
\eta^{(2)}(s)
\right\|_{\rm F}
}{
\sqrt{N_{\rm act}}
},
\end{equation}
where $N_{\rm act}$ is the number of active two-body matrix elements in the same representation used to evaluate the Frobenius norm. In a canonical antisymmetric pair basis with particle-particle/hole-hole support,
\begin{equation}
N_{\rm act}
=
2
\binom{A}{2}
\binom{N_{\rm sp}-A}{2},
\end{equation}
The accumulated intensive generator action is
\begin{equation}
\mathcal I_{\eta_2}
=
\int_0^{s_f}
{\rm d}s\,
\overline{\eta}_2(s).
\end{equation}
This quantity measures the cumulative strength with which an error in the reconstructed 3-RDM can enter the 2-RDM equation through the two-body generator. 

We next monitor whether the reconstructed RDMs retain their defining algebraic and particle-number properties~\cite{Akbari2012PRB,Lackner2015PRA}. The Hermiticity residuals are
\begin{equation}
\epsilon_{\rm H}^{(k)}
=
\max_s
\frac{
\left\|
\varrho^{(k)}(s)
-
\varrho^{(k)\dagger}(s)
\right\|_{\rm F}
}{
\left\|
\varrho^{(k)}(s)
\right\|_{\rm F}
},
\qquad
k=1,2,\qquad \epsilon_H = \max \left(\epsilon_H^{(1)}, \epsilon_H^{(2)}\right).
\label{eq:hermiticity_residuals}
\end{equation}
No explicit Hermitization is applied when evaluating these residuals, so they provide direct tests of the independently assembled BBGKY equations. The particle-number trace residuals are
\begin{align}
\epsilon_{{\rm Tr}_1}
&=
\max_s
\left|
{\rm Tr}\varrho^{(1)}(s)-A
\right|,
\epsilon_{{\rm Tr}_2}
=
\max_s
\left|
{\rm Tr}\varrho^{(2)}(s)-A(A-1)
\right|.
\end{align}
Compatibility between the two RDM levels is measured by
\begin{equation}
\epsilon_c
=
\max_s
\frac{
\left\|
{\rm Tr}_2\varrho^{(2)}_{12}(s)
-
(A-1)\varrho^{(1)}_1(s)
\right\|_{\rm F}
}{
(A-1)
\left\|
\varrho^{(1)}_1(s)
\right\|_{\rm F}
}.
\end{equation}
The trace residuals constrain only scalar particle numbers, whereas $\epsilon_c$ tests the complete one-body matrix obtained by contracting the 2-RDM. The eigenvalues of the 1-RDM define the natural occupations~\cite{Lowdin1955PR},
\begin{equation}
\left\{
n_i(s)
\right\}
=
\operatorname{eig}
\left[
\varrho^{(1)}(s)
\right].
\end{equation}
For an ensemble-$N$-representable fermionic 1-RDM, they satisfy~\cite{Coleman1963RMP}
\begin{equation}
0\leq n_i\leq1.
\end{equation}
We therefore record the worst values encountered along the reverse flow,
\begin{equation}
n_{\min}^{\rm flow}
=
\min_s n_{\min}(s),
\qquad
n_{\max}^{\rm flow}
=
\max_s n_{\max}(s).
\end{equation}
Violations beyond the numerical tolerance indicate a failure of ensemble 1-RDM representability. At the two-body level, we test the $P$, $Q$, and $G$ necessary $N$-representability conditions~\cite{Garrod1964JMP,DePrince2024WCMS}. In composite-index notation,
\begin{align}
P_{(pq),(rs)}
=
\left\langle
a_p^\dagger
a_q^\dagger
a_s
a_r
\right\rangle,
Q_{(pq),(rs)}
=
\left\langle
a_p
a_q
a_s^\dagger
a_r^\dagger
\right\rangle,
G_{(pq),(rs)}
=
\left\langle
a_p^\dagger
a_q
a_s^\dagger
a_r
\right\rangle.
\end{align}
For an $N$-representable 2-RDM,
\begin{equation}
P\succeq0,
\qquad
Q\succeq0,
\qquad
G\succeq0.
\end{equation}
We report the lowest eigenvalues encountered along the flow,
\begin{equation}
\lambda_{\min}^{P}
=
\min_s\lambda_{\min}[P(s)],
\qquad
\lambda_{\min}^{Q}
=
\min_s\lambda_{\min}[Q(s)],
\qquad
\lambda_{\min}^{G}
=
\min_s\lambda_{\min}[G(s)].
\end{equation}
The eigenvalues are evaluated from the Hermitian parts of the corresponding matrices, while the discarded anti-Hermitian components are monitored independently through Eq.~\eqref{eq:hermiticity_residuals}. Negative eigenvalues whose magnitudes exceed the numerical resolution identify violations of necessary $N$-representability conditions. Conversely, nonnegative $P$, $Q$, and $G$ spectra are necessary but not sufficient for full $N$-representability.

Finally, the magnitude of the contraction-consistent tangent projection is quantified by
\begin{equation}
\epsilon_{\rm proj}
=
\max_s
\frac{
\left\|
\Delta\dot{\varrho}^{(2)}(s)
\right\|_{\rm F}
}{
\left\|
\dot{\varrho}_{\rm raw}^{(2)}(s)
\right\|_{\rm F}
}.
\end{equation}
This quantity measures how strongly the unprojected closure attempts to drive the trajectory away from the contraction-consistent manifold. A small contraction residual accompanied by a large $\epsilon_{\rm proj}$ should therefore be interpreted as active correction by the projection rather than intrinsic contraction consistency of the closure.

\subsection{Exact-benchmark diagnostics}

For sufficiently small systems, exact diagonalization and fixed-$A$ Fock-space inversion provide additional diagnostics that directly separate different error sources. Exact diagonalization gives
\begin{equation}
E_{\rm exact},
\qquad
\varrho_{\rm exact}^{(1)},
\qquad
\varrho_{\rm exact}^{(2)}.
\end{equation}
These quantities define the physical ground-state reference against which the total IMSRG(2) and BBGKY errors are assessed. We also propagate the stored IMSRG(2) generator trajectory in the complete fixed-$A$ Fock space. Within the integration tolerance, this produces the exact state associated with the retained generator path and gives
\begin{equation}
E_{\rm Fock},
\qquad
\varrho_{\rm Fock}^{(1)}(s),
\qquad
\varrho_{\rm Fock}^{(2)}(s),
\qquad
\varrho_{\rm Fock}^{(3)}(s).
\end{equation}
The resulting state is not generally identical to the exact ground state because the stored generator trajectory was obtained from the truncated IMSRG(2) Hamiltonian evolution. For comparison, we also evolve the one- and two-body density operators separately within IMSRG(2)~\cite{Hergert2016PhysRep,Parzuchowski2017PRC}, obtaining
\begin{equation}
E_{\rm elem},
\qquad
\varrho_{\rm elem}^{(1)},
\qquad
\varrho_{\rm elem}^{(2)}.
\end{equation}
This conventional elementwise operator evolution introduces an operator-side IMSRG(2) truncation for each evolved matrix element. Comparison with the fixed-$A$ Fock inversion and the BBGKY reconstruction separates this additional operator truncation from the error already contained in the Hamiltonian trajectory.
The actual IMSRG(2) many-body error per particle is
\begin{equation}
\Delta E_{\rm MB}
=
\frac{
\left|
E_{\rm IMSRG(2)}
-
E_{\rm exact}
\right|
}{A}.
\end{equation}
Unlike $\Delta E_{\rm path}$, this quantity requires an exact reference and is available only in benchmark calculations. The final-state RDM errors relative to the exact ground state are
\begin{equation}
\epsilon_{\rm exact}^{(k)}
=
\frac{
\left\|
\varrho^{(k)}
-
\varrho_{\rm exact}^{(k)}
\right\|_{\rm F}
}{
\left\|
\varrho_{\rm exact}^{(k)}
\right\|_{\rm F}
},
\qquad
k=1,2.
\end{equation}
These errors contain the IMSRG(2) Hamiltonian-path error together with the additional reconstruction or operator-evolution error of the method used to obtain $\varrho^{(k)}$. To isolate the accuracy of the RDM reconstruction on the retained generator path, we compare it with the fixed-$A$ Fock-space inversion,
\begin{equation}
\epsilon_{{\rm path}}^{(k)}(s)
=
\frac{
\left\|
\varrho^{(k)}(s)
-
\varrho_{\rm Fock}^{(k)}(s)
\right\|_{\rm F}
}{
\left\|
\varrho_{\rm Fock}^{(k)}(s)
\right\|_{\rm F}
},
\qquad
k=1,2.
\label{eq:path_rdm_error}
\end{equation}
We report both the endpoint value and the maximum value over the complete reverse trajectory. Because the BBGKY and Fock-space calculations follow the same stored generator path, Eq.~\eqref{eq:path_rdm_error} directly measures the accumulated RDM reconstruction error on that path. The corresponding energy differences are
\begin{align}
\Delta E_{\rm F\text{-}I}
&=
\frac{
\left|
E_{\rm Fock}
-
E_{\rm IMSRG(2)}
\right|
}{A},
\\
\Delta E_{{\rm B}\text{-}\rm F}^{{\rm X}}
&=
\frac{
\left|
E_{\rm BBGKY(2)}^{X}
-
E_{\rm Fock}
\right|
}{A}.
\end{align}
The first quantity measures the inconsistency already present between the truncated Hamiltonian evolution and the exact state generated by its stored path. The second measures the additional energy discrepancy introduced by the BBGKY reconstruction on that path. To expose possible enhancement or cancellation between the two contributions, we retain their signs before taking absolute values:
\begin{align}
E_{\rm BBGKY(2)}^{X}
-
E_{\rm IMSRG(2)}
={}&
\left(
E_{\rm BBGKY(2)}^{X}
-
E_{\rm Fock}
\right)
\nonumber\\
&+
\left(
E_{\rm Fock}
-
E_{\rm IMSRG(2)}
\right).
\label{eq:signed_error_decomposition}
\end{align}
Equation~\eqref{eq:signed_error_decomposition} identifies cases in which the scalable diagnostic $\Delta E_{{\rm path}}^{\rm X}$ is reduced by an accidental cancellation between the reconstruction and Hamiltonian-path discrepancies. The exact fixed-$A$ propagation also provides the 3-RDM along the stored path. We therefore define the spectator-source discrepancy
\begin{equation}
\Xi_X(s)
=
\left\|
\mathcal S_{\eta^{(2)}(s)}
\left[
\varrho_{\rm rec}^{(3),X}(s)
-
\varrho_{\rm Fock}^{(3)}(s)
\right]
\right\|_{\rm F},
\end{equation}
where $\mathcal S_{\eta^{(2)}}$ denotes the spectator contribution generated by the two-body generator in the 2-RDM flow equation. Equivalently, $\Xi_X(s)$ is the norm of the instantaneous error induced in $\dot{\varrho}^{(2)}$ by replacing the exact path 3-RDM with the reconstructed one.
Because $\varrho_{\rm rec}^{(3),X}$ is constructed from the evolving BBGKY 1-RDM and 2-RDM, $\Xi_X$ contains both the local closure-functional error and the effect of the RDM error already accumulated earlier along the trajectory. It should therefore be interpreted as the actual three-body source discrepancy entering the approximate 2-RDM evolution.
\section{The Pairing-Plus-Particle-Hole (P3H) Model}
\label{sec:sm_p3h}

\subsection{Hamiltonian and benchmark protocol}

We use the pairing-plus-particle-hole (P3H) model~\cite{HjorthJensen2010JPG,Smith2025PRC} as an exactly
solvable benchmark in which both pair-conserving and pair-breaking
correlations can be varied independently.  
The Hamiltonian employed in the calculations is
\begin{align}
&H = H_{\mathrm{sp}}+H_{\mathrm{pair}}+H_{\mathrm{ph}},\\
&H_{\mathrm{sp}}
=\delta\sum_{p=1}^{\Omega}(p-1)
\sum_{\sigma=\pm}a^\dagger_{p\sigma}a_{p\sigma},\\
&H_{\mathrm{pair}}
=-\frac{g}{2}\sum_{p,q=1}^{\Omega}
a^\dagger_{p+}a^\dagger_{p-}a_{q-}a_{q+},\\
&H_{\mathrm{ph}}
=\frac{b}{2}\sum_{r=1}^{\Omega}
\sum_{\substack{p,q=1\\p\neq q}}^{\Omega}
\left(
a^\dagger_{p+}a^\dagger_{q-}a_{r-}a_{r+}
+\mathrm{H.c.}
\right).
\label{eq:sm_p3h_ph}
\end{align}
Here, $p$ labels doubly degenerate and equally spaced energy levels and $+$ and $-$ denote its two time-reversed states.  
The restriction $p\neq q$ in Eq.~\eqref{eq:sm_p3h_ph} removes the pair-conserving channel already contained in $H_{\mathrm{pair}}$.  The parameter $g$ controls coherent
pair scattering, whereas $b$ couples a correlated pair to two
time-reversed states belonging to different levels and therefore
introduces pair-breaking particle-hole excitations.  Both terms conserve the particle number and the spin projection. The benchmark contains $A=4$ particles in $\Omega=4$ doubly degenerate levels, corresponding to $N_{\mathrm{sp}}=8$ single-particle states.  We set $\delta=1\,\mathrm{MeV}$ and use the fixed reference determinant in
which the two lowest levels are occupied, instead of the self-consistent Hartree-Fock state.  The interaction strengths are
sampled on the grid
\begin{align}
g=-1.2,-1.1,\ldots,1.2\ \mathrm{MeV}, b=-0.25,-0.20,\ldots,0.25\ \mathrm{MeV},
\end{align}
with the noninteracting point $(g,b)=(0,0)$ omitted.  The resulting grid contains 274 interacting Hamiltonians.  For every Hamiltonian we perform independent IMSRG(2) evolutions with the White, arctangent and imaginary-time generators~\cite{White2002JCP,Hergert2016PhysRep,Smith2025PRC}.  
For convergence, we require the final off-diagonal Hamiltonian norm to be smaller than $10^{-6}$.  
At each accepted point we compare five density reconstructions.  Exact diagonalization gives $E_{\mathrm{FCI}}$ and the exact one- and two-body
RDMs.  Exact fixed-$A$ inversion of the stored IMSRG(2) generator path
defines $E_{\mathrm{Fock}}$ and
$\varrho^{(k)}_{\mathrm{Fock}}$ for that same truncated trajectory.  These
quantities isolate the error of the IMSRG(2) path from the additional
error introduced when it is represented in reduced-density space.  We
then compare conventional elementwise operator evolution with the V, NY,
and TS closures of IMSRG(2)-BBGKY(2).

\begin{figure*}[h]
    \centering
    \includegraphics[width=\textwidth]{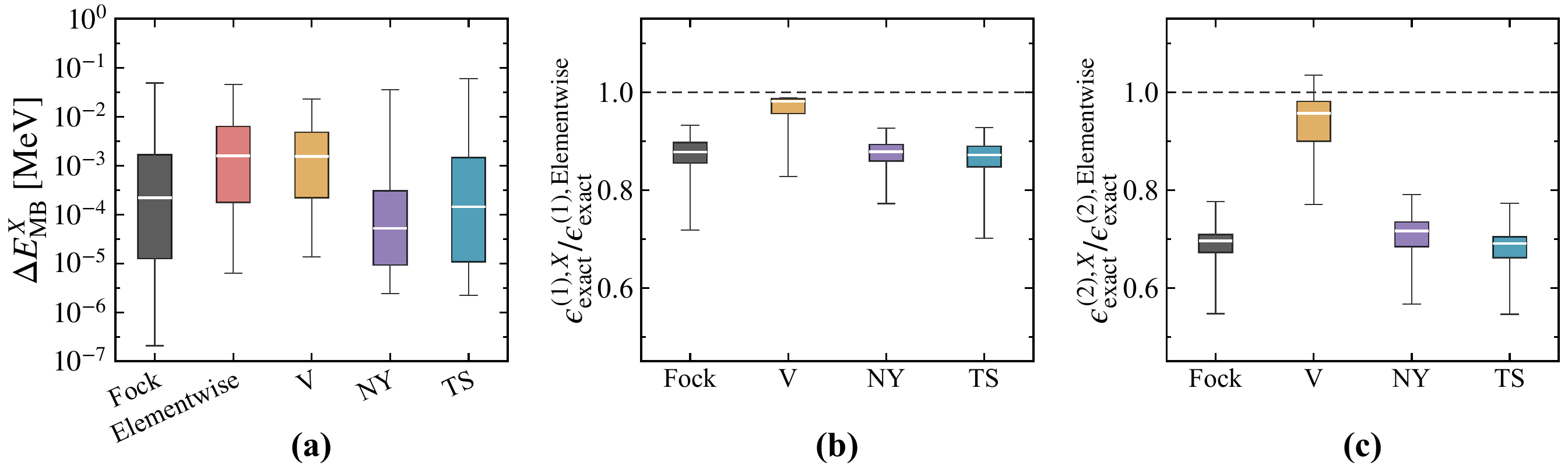}
\caption{Accuracy of the reconstructed observables for the 274 P3H Hamiltonians.  For each Hamiltonian and method, the errors are first averaged over the White, arctangent, and imaginary-time generators.  (a) Absolute many-body energy error per particle.  (b,c) Pointwise ratios of the relative 1- and 2-RDM errors, $\epsilon_{\rm exact}^{(k),X}/ \epsilon_{\rm exact}^{(k),{\rm Elementwise}}$, for $k=1$ and $2$, respectively.  Each ratio compares methods for the same Hamiltonian.
The dashed line denotes unity, and values below it indicate an
improvement over elementwise evolution.  Fock denotes exact
fixed-$A$ inversion of the corresponding stored IMSRG(2) trajectory.
Boxes span the interquartile range, white lines indicate the medians, and whiskers show the 5th and 95th percentiles.}
\label{fig:smc1}
\end{figure*}

\subsection{Accuracy of the reconstructed densities}

Fig.~\ref{fig:smc1} summarizes the errors with respect to FCI over the
full 274-point sample.  For each Hamiltonian and method, the errors are
first averaged over the three generators.  As in the preceding section,
the energy error is quoted per particle, while the RDM errors are
normalized Frobenius distances.
Here and after, we use available FCI results as exact benchmarks. Panel (a) shows the distributions of $\Delta E_{\rm MB}^{X}$.  
The NY closure gives the smallest median energy error,
$5.2\times10^{-5}\,\mathrm{MeV}$ per particle, compared with
$1.6\times10^{-3}\,\mathrm{MeV}$ for elementwise evolution.  
To expose the paired improvement of the reconstructed RDMs, panels (b) and (c) show the pointwise ratios
$\epsilon_{\rm exact}^{(k),X}/
\epsilon_{\rm exact}^{(k),\mathrm{Elementwise}}$ for the same
Hamiltonian.  Values below unity therefore indicate an improvement over elementwise evolution.  For the 1-RDM, the median ratios are $0.879$ for NY and $0.872$ for TS, while for the 2-RDM they are $0.717$ and $0.691$, respectively.  Both closures improve upon elementwise evolution at all 274 points.  These ratios closely track the corresponding exact Fock-inversion values, $0.879$ for the 1-RDM and $0.697$ for the 2-RDM.
Thus, over most of the grid, the remaining RDM error is controlled primarily by the stored IMSRG(2) trajectory rather than by the BBGKY(2) reconstruction.

\begin{figure*}[t]
    \centering
    \includegraphics[width=0.9\textwidth]{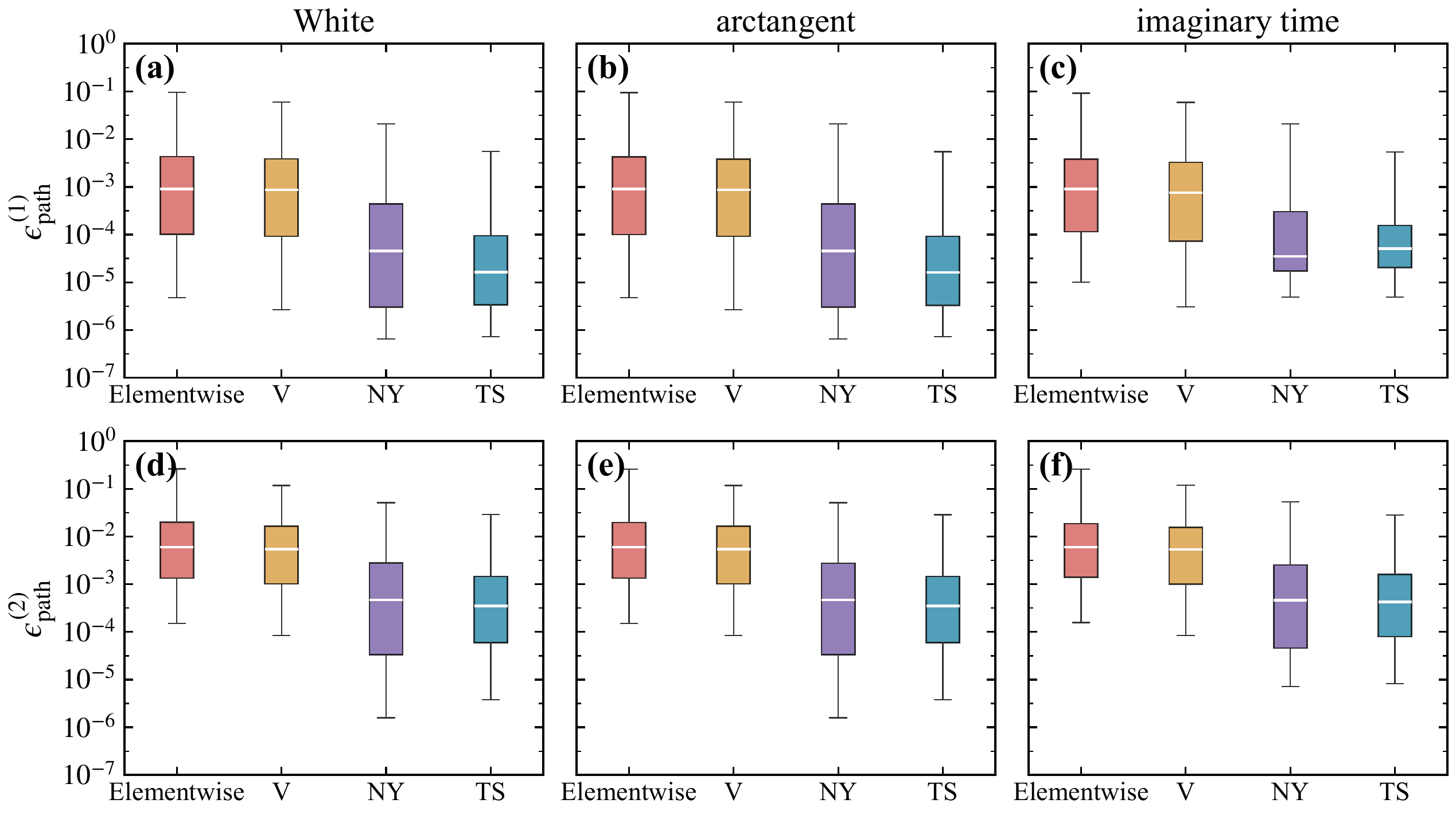}
    \caption{Reconstruction errors relative to exact fixed-$A$ inversion
    of the same stored IMSRG(2) path.  The upper and lower rows show the
    1-RDM and 2-RDM errors, respectively, while the columns correspond to
    the White, arctangent, and imaginary-time generators.  Box and
    whisker conventions are the same as in Fig.~\ref{fig:smc1}.  NY and
    TS improve the path-matched 2-RDM by approximately one order of
    magnitude for each generator separately.}
    \label{fig:smc2}
\end{figure*}

To remove the intrinsic IMSRG(2) truncation error from this comparison, we next use the exact fixed-$A$ inversion of the same stored generator trajectory as the reference.
Fig.~\ref{fig:smc2} displays the quantity defined in Eq.~(\ref{eq:path_rdm_error}) separately for the three
generators.  The hierarchy of the reconstruction schemes is essentially generator independent.  
For the 2-RDM, the median errors are
$(5.99$--$6.00)\times10^{-3}$ for elementwise evolution,
$(5.33$--$5.39)\times10^{-3}$ for V,
$(4.63$--$4.71)\times10^{-4}$ for NY, and
$(3.49$--$4.26)\times10^{-4}$ for TS.  A similar improvement is obtained for the 1-RDM.  The agreement among the three columns also shows that the gain from NY and TS is not produced by averaging mutually compensating generator errors.

\begin{figure*}[b]
    \centering
    \includegraphics[width=\textwidth]{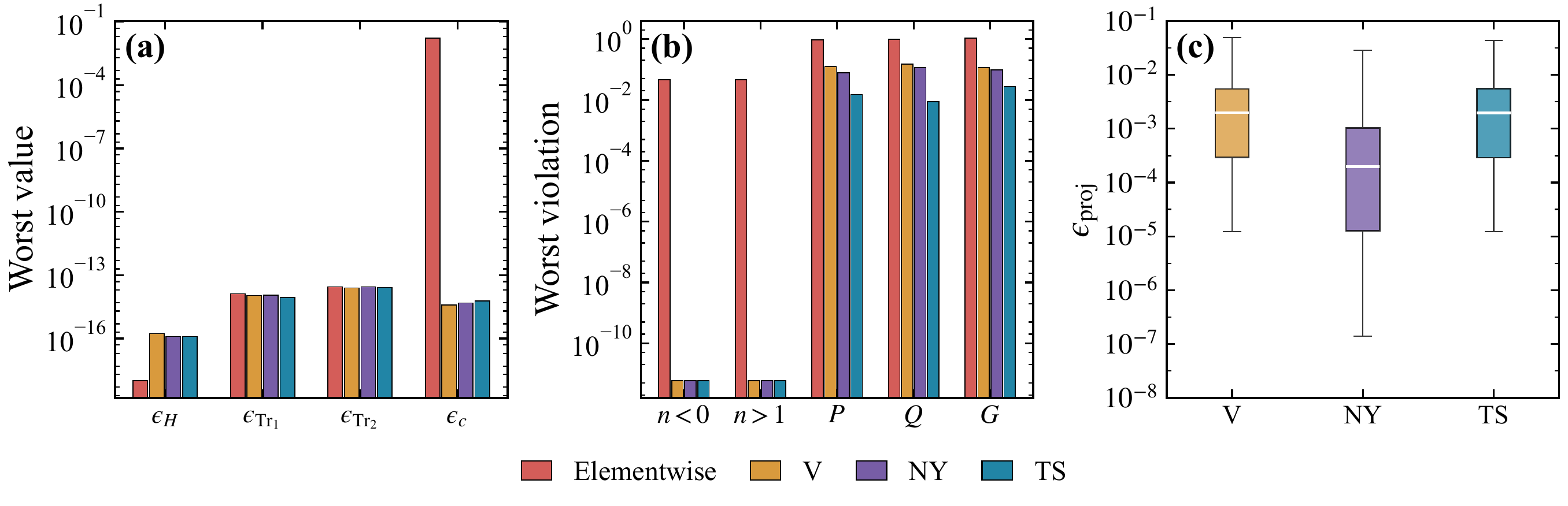}
    \caption{Constraint and necessary $N$-representability diagnostics
    over the converged P3H ensemble.  (a) Worst Hermiticity, one-body
    trace, two-body trace, and contraction residuals.  (b) Worst
    violation of the natural-occupation bounds and the $P$, $Q$, and $G$
    conditions.  A zero value denotes no violation; displayed bars are
    violation magnitudes.  (c) Distribution of the largest projection
    correction encountered along each reverse flow.  The boxes span the
    interquartile range and the whiskers show the 5th and 95th
    percentiles.}
    \label{fig:smc3}
\end{figure*}

\subsection{Flow constraints, strong-coupling behavior, and signed energy errors}

Fig.~\ref{fig:smc3} collects the algebraic and necessary
$N$-representability diagnostics defined above.  For each Hamiltonian
and generator, every diagnostic is extremized over the corresponding
complete flow: the forward operator flow for elementwise evolution and
the reverse RDM flow for the three BBGKY closures.  Panels (a) and (b)
then report the worst value over the complete 274-point sample.  Exact Fock-space inversion, which preserves these conditions to numerical precision, is not shown.  
Hermiticity and the one- and two-body traces remain at numerical precision for all displayed methods, with largest
residuals of $1.7\times10^{-16}$, $1.3\times10^{-14}$, and
$2.8\times10^{-14}$, respectively.  The contraction projection keeps the BBGKY contraction residual below $5.9\times10^{-15}$, whereas the elementwise result reaches $1.7\times10^{-2}$.  Violations of the natural-occupation interval $[0,1]$ remain below $5.9\times10^{-12}$ for the BBGKY replays and $4.6\times10^{-2}$ for elementwise evolutions. The largest violations of these properties in elementwise evolutions arise in the strongest coupling cases, emphasizing the advantage of this new replay approach.

The $P$, $Q$, and $G$ conditions are more sensitive.  
In panel (b), we plot the violation magnitudes $\max(0,-\lambda_{\min}^{P,Q,G})$.  
The largest violation among these three conditions is approximately $1.07$ for elementwise evolution, compared with $0.151$ for V, $0.114$ for NY, and $0.0275$ for TS.  
Thus, although none of the approximate RDM flows strictly enforces two-positivity, all three BBGKY closures improve upon
elementwise evolution, with TS providing the strongest suppression.
These negative eigenvalues quantify a limitation of the approximate RDM reconstruction rather than a loss of Hermiticity or particle-number normalization.  
Panel (c) shows the contraction-projection correction,
which is defined only for the BBGKY flows.  Its median
$\epsilon_{\mathrm{proj}}$ is $2.0\times10^{-3}$ for V,
$2.0\times10^{-4}$ for NY, and $2.0\times10^{-3}$ for TS, while the strongest-coupling tail extends to approximately $0.1$.

\begin{figure*}[t]
    \centering
    \includegraphics[width=0.8\textwidth]{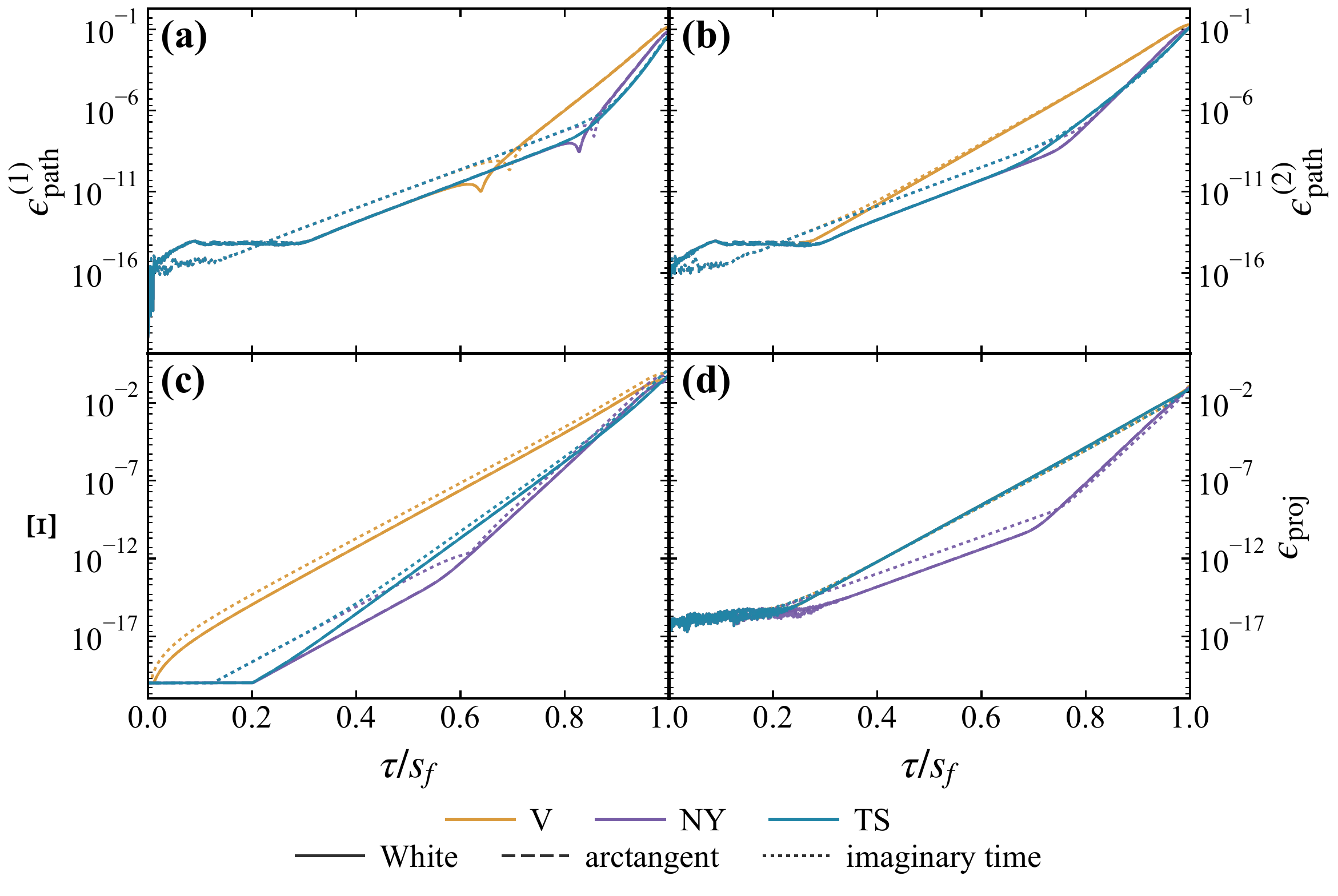}
    \caption{Flow-resolved diagnostics for the strong-coupling point
    $(g,b)=(1.2,0.25)\,\mathrm{MeV}$ recomputed with $s_{\max}=120$.
    (a) 1-RDM and (b) 2-RDM reconstruction errors relative to exact Fock
    inversion at the same reverse-flow node.  (c) Spectator-source
    discrepancy $\Xi$.  (d) Projection correction.  Colors distinguish
    the V, NY, and TS closures, and line styles distinguish the three
    generators.}
    \label{fig:smc4}
\end{figure*}

The strongest coupling case in our calculations is illustrated in Fig.~\ref{fig:smc4}.  
We introduce the normalized reverse coordinate $\tau/s_f$, which is zero at the decoupled endpoint and unity at the original Hamiltonian.  
All three generators produce smooth and mutually consistent growth of the reconstruction errors.  
At the physical endpoint, the generator-averaged 2-RDM path errors are $0.203$ for V, $0.131$ for NY, and $0.130$ for TS.  The corresponding 1-RDM errors are $0.156$, $0.084$, and $0.048$, respectively.  
This point therefore lies near the quantitative boundary of the two-body closures, but its behavior remains controlled rather than becoming numerically unstable.  
The spectator-source discrepancy $\Xi$ and the projection
correction grow in the same region of the reverse path, providing useful internal warnings when the reconstruction becomes difficult.

Finally, exact Fock inversion permits a signed decomposition of the energy-consistency defect. We define the common IMSRG-path contribution and the reconstruction contributions as
\begin{align}
d_{\mathrm{F-I}}
&=\frac{E_{\mathrm{Fock}}-E_{\mathrm{IMSRG(2)}}}{A},
d_{\mathrm{E-F}}
=\frac{E_{\mathrm{Elem}}-E_{\mathrm{Fock}}}{A},\\
d_{\mathrm{B-F}}^{X}
&=\frac{E_{\mathrm{BBGKY(2)}}^{X}-E_{\mathrm{Fock}}}{A},
\qquad X\in\{\mathrm{V},\mathrm{NY},\mathrm{TS}\}.
\end{align}
The corresponding total signed defects are
\begin{align}
d_{\mathrm{E-I}}
&=d_{\mathrm{F-I}}+d_{\mathrm{E-F}}
=\frac{E_{\mathrm{Elem}}-E_{\mathrm{IMSRG(2)}}}{A},\\
d_{\mathrm{B-I}}^{X}
&=d_{\mathrm{F-I}}+d_{\mathrm{B-F}}^{X}
=\frac{E_{\mathrm{BBGKY(2)}}^{X}
-E_{\mathrm{IMSRG(2)}}}{A}.
\end{align}
\begin{figure*}[t]
    \centering
    \includegraphics[width=0.8\textwidth]{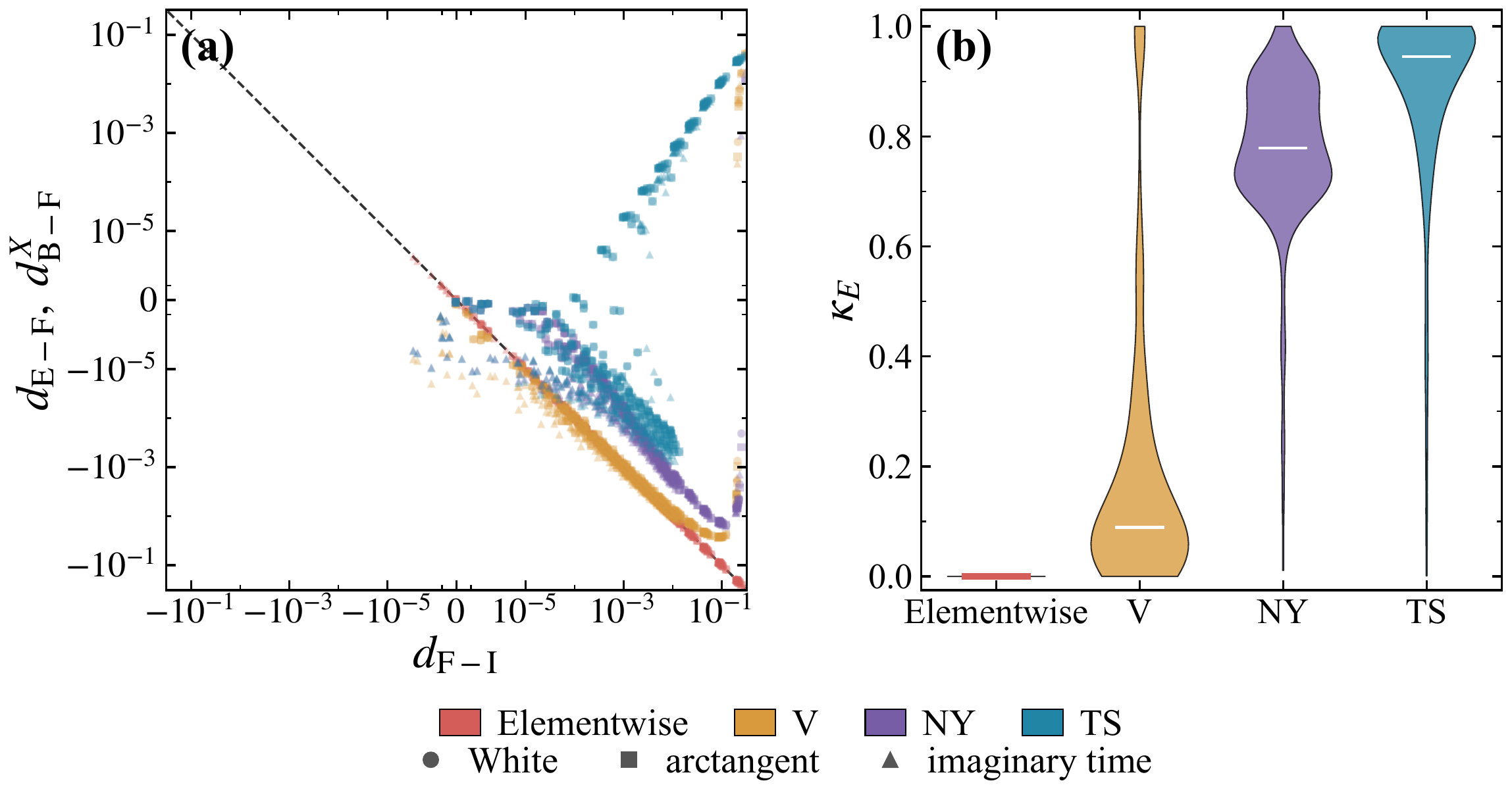}
\caption{Signed decomposition of the P3H energy-consistency defect.
(a) Signed reconstruction contributions relative to exact Fock inversion
as functions of the IMSRG-path contribution $d_{\mathrm{F-I}}$.
Elementwise evolution is represented by $d_{\mathrm{E-F}}$, while closure
$X\in\{\mathrm{V},\mathrm{NY},\mathrm{TS}\}$ is represented by
$d_{\mathrm{B-F}}^{X}$. The dashed line denotes
$d_{\mathrm{E-F}}=-d_{\mathrm{F-I}}$ or
$d_{\mathrm{B-F}}^{X}=-d_{\mathrm{F-I}}$, corresponding to exact
cancellation. Colors distinguish the reconstruction methods and marker
shapes distinguish the generators.
(b) Distributions of the cancellation factor $\kappa_E$ for elementwise
evolution and the V, NY, and TS closures. Horizontal segments indicate
the corresponding medians.}
    \label{fig:smc5}
\end{figure*}
Here, $d_{\mathrm{F-I}}$ measures the difference between the stored IMSRG(2) trajectory and its exact fixed-$A$ inversion, whereas $d_{\mathrm{E-F}}$ and $d_{\mathrm{B-F}}^{X}$ measure the additional differences introduced by elementwise evolution and BBGKY reconstruction,
respectively. The path diagnostic used in the main text is
$\Delta E_{\mathrm{path}}^{X}=|d_{\mathrm{B-I}}^{X}|$.
To quantify cancellation between the two signed contributions, we define
\begin{align}
\kappa_E^{\mathrm{Elem}}
=
\frac{|d_{\mathrm{E-I}}|}
{|d_{\mathrm{F-I}}|+|d_{\mathrm{E-F}}|},
\kappa_E^{X}
=
\frac{|d_{\mathrm{B-I}}^{X}|}
{|d_{\mathrm{F-I}}|+|d_{\mathrm{B-F}}^{X}|},
\qquad 0\leq\kappa_E\leq1.
\end{align}
Perfect cancellation gives $\kappa_E=0$, whereas $\kappa_E\simeq1$
indicates little cancellation or mutually reinforcing contributions.

Fig.~\ref{fig:smc5} shows that the cancellation pattern depends
strongly on the reconstruction method. The median $\kappa_E$ is
approximately $1.2\times10^{-13}$ for elementwise evolution, $0.089$ for V, $0.779$ for NY, and $0.945$ for TS. Moreover, $52.6\%$ of the V results have $\kappa_E<0.1$, compared with less than $1\%$ for NY or TS. 
A small V energy defect can therefore result from cancellation between two much larger signed contributions, whereas the TS defect is considerably less susceptible to this failure mode. This behavior motivates the use of the TS reconstruction in the energy-error calibration of the main text.
The near-vanishing elementwise value has a different origin. Because the projected operator flow is linear, contracting the complete set of elementwise-evolved density operators with the initial Hamiltonian reproduces the IMSRG(2) energy up to numerical integration errors.
Consequently,
$d_{\mathrm{E-F}}\simeq-d_{\mathrm{F-I}}$ and
$E_{\mathrm{Elem}}\simeq E_{\mathrm{IMSRG(2)}}$ by construction. This energy agreement does not imply that the reconstructed RDM is accurate. The energy probes only one particular contraction of its matrix elements. Errors in other directions of RDM space can remain large and may be sampled much more strongly by other one- and two-body observables. Thus, elementwise evolution can reproduce the IMSRG(2) energy while giving appreciably different predictions for observables with different operator structure.

\section{Hamiltonian and model-space setups for the path diagnostic}

This section documents the Hamiltonians, model spaces, coupling grids, and flow diagnostics underlying Fig.~3 of the main text. The complete dataset contains 1488 toy-model calculations and 126 distinct nuclear-Hamiltonian calculations. Exact diagonalization provides the toy-model benchmarks, while the neutron-matter benchmarks are obtained from either sector-exact diagonalization or independent FCIQMC calculations~\cite{Booth2009JCP,Hu2026PRL}. 
Exact information is used to select and validate the benchmark sample and to evaluate $d_{\rm MB}$ and it is not required to obtain $d_{\rm path}$. 
All calculations included here use the arctangent generator and the TS closure.
For the toy models, the forward flow is integrated with an adaptive midpoint step-doubling algorithm using initial, maximum, and minimum steps of $0.1$, $0.2$, and $10^{-7}$, respectively, with relative and absolute tolerances of $10^{-6}$ and $10^{-9}$. The flow is regarded as converged when the off-diagonal Hamiltonian norm falls below $10^{-6}$ MeV. The accepted generator trajectory is stored and replayed backward using BBGKY(2). This local integration-error control is distinct from the condition used to terminate the forward flow.
Two operational criteria restrict the sampled coupling domain. First, a trial is rejected if the off-diagonal norm increases along the accepted IMSRG(2) flow beyond roundoff. Second, for a nonzero exact correlation energy, we require
\begin{align}
\frac{|E_{\rm IMSRG(2)}-E_{\rm HF}|}
{|E_{\rm exact}-E_{\rm HF}|}<2,
\end{align}
where $E_{\rm HF}$ denotes the energy of the reference determinant used in the calculation. Reference-sector checks additionally exclude points for which a ground-state crossing changes the target branch. 
These criteria are set to cover both strongly and weakly coupled points in our numerical experiments.

\subsection{Hamiltonians and model spaces}

The four model spaces used in toy model calculations are
\begin{align}
(A,L,N_{\rm sp})
\in
\{(2,2,4),(2,4,8),(4,4,8),(4,6,12)\},
\qquad N_{\rm sp}=2L.
\end{align}
For the spinful models, $L$ counts spatial sites or orbitals with two spin states. The spinless models use two identical uncoupled flavors, whereas the LMG model has two levels with degeneracy $L$ each. The one-body Hamiltonian is diagonalized before normal ordering, and its lowest $A$ eigenstates define the reference determinant. The lattice chains below have open boundaries. The site and orbital indices start at zero, so the first alternating hopping is proportional to $1+\delta$. Numerical parameters are expressed in the energy units of the corresponding model. 

For the four spinful chains, $n_{i\sigma}=c^\dagger_{i\sigma}c_{i\sigma}$ and $n_i=n_{i\uparrow}+n_{i\downarrow}$. The Hubbard Hamiltonian~\cite{Hubbard1963PRSA} is
\begin{align}
H_{\rm Hubbard}
={}&
-\sum_{i=0}^{L-2}\sum_\sigma
\left(c^\dagger_{i\sigma}c_{i+1,\sigma}+{\rm H.c.}\right)
+
U\sum_{i=0}^{L-1}n_{i\uparrow}n_{i\downarrow}.
\end{align}
The extended-Hubbard Hamiltonian~\cite{Nakamura2000PRB} includes a fixed nearest-neighbor interaction:
\begin{align}
H_{\rm extended}
=
-\sum_{i=0}^{L-2}\sum_\sigma
\left(c^\dagger_{i\sigma}c_{i+1,\sigma}+{\rm H.c.}\right)
+
U\sum_{i=0}^{L-1}n_{i\uparrow}n_{i\downarrow}
+
0.40\sum_{i=0}^{L-2}n_i n_{i+1}.
\end{align}
The ionic-Hubbard Hamiltonian~\cite{Fabrizio1999PRL} contains a staggered one-body potential:
\begin{align}
H_{\rm ionic}
=
-\sum_{i=0}^{L-2}\sum_\sigma
\left(c^\dagger_{i\sigma}c_{i+1,\sigma}+{\rm H.c.}\right)
+
U\sum_{i=0}^{L-1}n_{i\uparrow}n_{i\downarrow}
+
0.50\sum_{i=0}^{L-1}(-1)^i n_i.
\end{align}
The SSH-Hubbard Hamiltonian~\cite{Su1979PRL,Hubbard1963PRSA} instead has alternating hopping amplitudes:
\begin{align}
H_{\rm SSH\text{-}Hubbard}
=
-\sum_{i=0}^{L-2}\sum_\sigma
\left[1+0.35(-1)^i\right]
\left(c^\dagger_{i\sigma}c_{i+1,\sigma}+{\rm H.c.}\right)
+
U\sum_{i=0}^{L-1}n_{i\uparrow}n_{i\downarrow}.
\end{align}
Only $U$ is scanned in these four families.

For the four spinless families, $\tau=1,2$ labels uncoupled flavors and $n_{i\tau}=c^\dagger_{i\tau}c_{i\tau}$. The spinless $t$-$V$ Hamiltonian~\cite{Rigol2009PRA} is
\begin{align}
H_{t\text{-}V}
=
\sum_{\tau=1}^{2}
\bigg[
&-\sum_{i=0}^{L-2}
\left(c^\dagger_{i\tau}c_{i+1,\tau}+{\rm H.c.}\right)
+
V_1\sum_{i=0}^{L-2}n_{i\tau}n_{i+1,\tau}
\bigg].
\end{align}

The spinless SSH-$V$ Hamiltonian~\cite{Su1979PRL,Rigol2009PRA} is
\begin{align}
H_{\rm SSH\text{-}V}
=
\sum_{\tau=1}^{2}
\bigg[
-\sum_{i=0}^{L-2}
\left[1+0.35(-1)^i\right]
\left(c^\dagger_{i\tau}c_{i+1,\tau}+{\rm H.c.}\right)
+
V_1\sum_{i=0}^{L-2}n_{i\tau}n_{i+1,\tau}
\bigg].
\end{align}

The spinless ionic-$V$ Hamiltonian~\cite{Rice1982PRL,Rigol2009PRA} is
\begin{align}
H_{\rm ionic\text{-}V}
=
\sum_{\tau=1}^{2}
\bigg[
-\sum_{i=0}^{L-2}
\left(c^\dagger_{i\tau}c_{i+1,\tau}+{\rm H.c.}\right)
+
0.50\sum_{i=0}^{L-1}(-1)^i n_{i\tau}
+
V_1\sum_{i=0}^{L-2}n_{i\tau}n_{i+1,\tau}
\bigg].
\end{align}

The frustrated spinless Hamiltonian~\cite{Rigol2009PRA} additionally includes next-nearest-neighbor hopping and interactions:
\begin{align}
H_{\rm frustrated}
=
\sum_{\tau=1}^{2}
\bigg[
-\sum_{i=0}^{L-2}
\left(c^\dagger_{i\tau}c_{i+1,\tau}+{\rm H.c.}\right)
-
0.35\sum_{i=0}^{L-3}
\left(c^\dagger_{i\tau}c_{i+2,\tau}+{\rm H.c.}\right)+
V_1\sum_{i=0}^{L-2}n_{i\tau}n_{i+1,\tau}
+
0.40\sum_{i=0}^{L-3}n_{i\tau}n_{i+2,\tau}
\bigg].
\end{align}
Here $V_1$ is the scanned coupling in all four families, and there is no interaction between flavors. Sums with an empty index range vanish.

For the Lipkin--Meshkov--Glick model~\cite{Lipkin1965NP}, let $a^\dagger_{\pm m}$ create a fermion in the upper or lower level, with $m=0,\ldots,L-1$. We use
\begin{align}
H_{\rm LMG}
=
J_z+\frac{\chi}{2(L-1)}
\left(J_+^2+J_-^2\right), 
J_z
=
\frac{1}{2}\sum_{m=0}^{L-1}
\left(a^\dagger_{+m}a_{+m}-a^\dagger_{-m}a_{-m}\right),
J_+
=
\sum_{m=0}^{L-1}a^\dagger_{+m}a_{-m},
J_-=J_+^\dagger.
\end{align}
The level separation is set as $1.0$ MeV, and $\chi$ is varied.

The unequal-level pairing Hamiltonian~\cite{Richardson1963PL} is
\begin{align}
H_{\rm unequal}
&=
\sum_{p=0}^{L-1}\sum_\sigma
\epsilon_p n_{p\sigma}
-\frac{G}{2}\sum_{p,q=0}^{L-1}P_p^\dagger P_q,
\\
(\epsilon_0,\epsilon_1,\epsilon_2,\epsilon_3)
&=(0,0.55,1.70,3.20),
\qquad L=4,
\\
\epsilon_p
&=
\left(\frac{3.2p}{L-1}\right)^{1.15},
\qquad L\neq4.
\end{align}
The pair operators have the same definition as in the P3H model, and $G$ is scanned.

The Anderson-impurity model~\cite{Anderson1961PR} consists of one interacting orbital $d$ and $L-1$ bath orbitals $b_m$:
\begin{align}
H_{\rm impurity}
={}&
\sum_{m=0}^{L-2}\sum_\sigma
\epsilon_m b^\dagger_{m\sigma}b_{m\sigma}
-\frac{U}{2}\sum_\sigma n_{d\sigma}
+
U n_{d\uparrow}n_{d\downarrow}
+
0.45\sum_{m=0}^{L-2}\sum_\sigma
\left(d^\dagger_\sigma b_{m\sigma}+{\rm H.c.}\right),
\\
\epsilon_m
={}&
-1.5+\frac{3m}{L-2},
\qquad L>2.
\end{align}
For $L=2$, the single bath energy is $-1.5$. The impurity energy is $-U/2$, so the one-body eigenbasis is recomputed as $U$ varies.

The finite-chain realization of the periodic Anderson model~\cite{Guerrero1996PRB} has $L/2$ sites, each with a conduction orbital $c$ and a localized orbital $f$:
\begin{align}
H_{\rm periodic\ Anderson}
=
-\sum_{j=0}^{L/2-2}\sum_\sigma
\left(c^\dagger_{j\sigma}c_{j+1,\sigma}+{\rm H.c.}\right)
-\frac{U}{2}\sum_{j=0}^{L/2-1}\sum_\sigma n^f_{j\sigma}+
0.50\sum_{j=0}^{L/2-1}\sum_\sigma
\left(c^\dagger_{j\sigma}f_{j\sigma}+{\rm H.c.}\right)
+
U\sum_{j=0}^{L/2-1}n^f_{j\uparrow}n^f_{j\downarrow}.
\end{align}
Here $n^f_{j\sigma}=f^\dagger_{j\sigma}f_{j\sigma}$. Despite the conventional model name, the finite chain used here has open boundaries. The scanned coupling is $U$.

The Kanamori--Hund model~\cite{Kanamori1963PTP,Georges2013ARCMP} also contains $L/2$ sites, now with two interacting orbitals $a=0,1$ per site. With $n_{ja\sigma}=c^\dagger_{ja\sigma}c_{ja\sigma}$, $S^+_{ja}=c^\dagger_{ja\uparrow}c_{ja\downarrow}$, and $P^\dagger_{ja}=c^\dagger_{ja\uparrow}c^\dagger_{ja\downarrow}$, its Hamiltonian is
\begin{align}
H_{\rm KH}
={}&
-0.70\sum_{j=0}^{L/2-2}\sum_{a,\sigma}
\left(c^\dagger_{ja\sigma}c_{j+1,a,\sigma}+{\rm H.c.}\right)
+
0.20\sum_{j,\sigma}
\left(n_{j1\sigma}-n_{j0\sigma}\right)
\nonumber\\
&+
U\sum_{j,a}n_{ja\uparrow}n_{ja\downarrow}
+
U'\sum_j
\left(n_{j0\uparrow}n_{j1\downarrow}
+n_{j0\downarrow}n_{j1\uparrow}\right)
\nonumber\\
&+
(U'-J)\sum_{j,\sigma}n_{j0\sigma}n_{j1\sigma}
-
J\sum_j
\left(S^+_{j0}S^-_{j1}+{\rm H.c.}\right)
+
J\sum_j
\left(P^\dagger_{j0}P_{j1}+{\rm H.c.}\right),
\\
U'={}&U-2J.
\end{align}
All unrestricted site sums run over $j=0,\ldots,L/2-1$. The stored calculations use two parameterizations of the scanned variable $x$:
\begin{align}
(U,J)
=
\begin{cases}
(2,x), & (A,L)=(2,2),(4,4),\\
(2x,0.4x), & (A,L)=(2,4),(4,6).
\end{cases}
\end{align}
Thus, the first parameterization varies the Hund coupling at fixed intraorbital repulsion, whereas the second scales the complete interaction at fixed $J/U=0.2$. These choices are stated separately because they define different coupling paths within the same Hamiltonian family.

The trapped Gaudin--Yang Hamiltonian~\cite{Gaudin1967PLA,Yang1967PRL,Grining2015NJP} is represented in the lowest $L$ one-dimensional harmonic-oscillator orbitals:
\begin{align}
H_{\rm GY}
={}&
\sum_{p=0}^{L-1}\sum_\sigma
\left(p+\frac{1}{2}\right)n_{p\sigma}
+
g\sum_{p,q,r,s=0}^{L-1}
I_{pqrs}\,
c^\dagger_{p\uparrow}c^\dagger_{q\downarrow}
c_{s\downarrow}c_{r\uparrow},
\\
I_{pqrs}
={}&
\int_{-\infty}^{\infty}
\phi_p(x)\phi_q(x)\phi_r(x)\phi_s(x)\,dx.
\end{align}
The real oscillator functions $\phi_p$ are normalized with $\hbar=m=\omega=1$. Both attractive and repulsive values of the contact coupling $g$ are included.

Finally, the quadrupole--quadrupole toy model uses a separable, spin-independent one-body matrix:
\begin{align}
H_{\rm QQ}
&=
\sum_{p=0}^{L-1}\sum_\sigma p\,n_{p\sigma}
-\frac{\chi}{2}:\!Q^2\!:_{\rm vac},
\\
Q
&=
\sum_{p,q=0}^{L-1}\sum_\sigma
q_{pq}c^\dagger_{p\sigma}c_{q\sigma},
\\
q_{pp}
&=
p+\frac{1}{2}-\frac{L}{2},
\qquad
q_{p,p+2}=q_{p+2,p}
=
\frac{1}{2}\sqrt{(p+1)(p+2)}.
\end{align}
All other matrix elements vanish. Particle-vacuum normal ordering in $:\!Q^2\!:_{\rm vac}$ retains only the two-body part, and $\chi$ is scanned. 

For every one-parameter family, the coupling grid is constructed from
\begin{align}
f_i^{(n)}
=
\left[
0.08+0.92\frac{i-1}{n-1}
\right]^{1.35},
\qquad i=1,\ldots,n.
\label{eq:sm_coupling_fraction}
\end{align}
For a one-sided grid, the 25 points are $x_i=x_+f_i^{(25)}$. For a two-sided grid, they are
\begin{align}
\left\{-|x_-|f_i^{(12)}\right\}_{i=1}^{12}
\cup
\left\{x_+f_i^{(13)}\right\}_{i=1}^{13}.
\label{eq:sm_two_sided_grid}
\end{align}
Equations~(\ref{eq:sm_coupling_fraction}) and (\ref{eq:sm_two_sided_grid}), together with the endpoints in Table~\ref{tab:sm_toy_endpoints}, specify all one-parameter coupling points. The scanned variable $x$ denotes $U$ for the spinful chains and Anderson models, $V_1$ for the spinless chains, $G$ for unequal pairing, $g$ for Gaudin--Yang, and $\chi$ for LMG and quadrupole--quadrupole. Full-precision couplings are retained in the accompanying data.

Production grids normally extend to $90\%$ of the accepted scan boundary in the first three spaces. For the $A=4,L=6$ non-P3H calculations, the fraction is reduced to $70\%$. When no failure is found before the prescribed scan limit, that limit is used instead of an inferred instability boundary. Several P3H points are moved inward along their coupling rays to satisfy the acceptance criteria. The $A=4,L=6$ quadrupole--quadrupole grid has a target-state crossing between points 13 and 14. Only the first 13 points of its nominal 25-point grid are retained, without rescaling the grid.
The resulting toy-model sample contains
\begin{align}
400+300+400+388=1488
\end{align}
calculations in the $(A,L)=(2,2),(2,4),(4,4),(4,6)$ spaces, respectively. The four spinless families are not included in the $A=2,L=4$ space. Every other family-space combination contains 25 points, except for the 13 retained quadrupole--quadrupole points specified above.

For neutron matter, Table~\ref{tab:sm_path_pnm_interactions} lists 22 NN-only or NN-plus-three-nucleon Hamiltonian combinations. Three-nucleon contributions, when present, are included in the normal-ordered two-body approximation. The sector-exact benchmark set contains all 22 combinations at
\begin{align}
\rho=0.04,\ 0.08,\ 0.12,\ 0.16,\ 0.20~{\rm fm}^{-3}
\end{align}
in the $(A,N_{\rm sp})=(14,38)$ space, giving 110 calculations. For Minnesota, diagonalization uses the $\boldsymbol K=0$, $M_s=0$, spatial $O_h$ $A_{1g}$ sector of dimension $286917$. For the other interactions, it uses the $\boldsymbol K=0$ sector transforming as $A_{1g}$ under the program's combined spatial-spin $D_{4h}$ operations, of dimension $3412113$. The largest directly recomputed Ritz residual is $9.82\times10^{-9}$ MeV. These energies are exact within the specified sectors.
Additional FCIQMC benchmarks are available for N$^3$LO Texas 394 with its nonlocal three-nucleon interaction and N$^2$LO EMN 500 with its nonlocal three-nucleon interaction. They cover
\begin{align}
(A,N_{\rm sp})
&=(14,38),\ (14,54),\ (38,54),
\\
\rho
&=0.08,\ 0.12,\ 0.16,\ 0.20~{\rm fm}^{-3}.
\end{align}
These are the spaces denoted $1/2$, $1/3$, and $2/3$, respectively. Of the 24 FCIQMC comparisons, eight correspond to Hamiltonian-space-density combinations already included in the sector-exact set. We retain the sector-exact entry for each of these eight combinations and do not count the alternative FCIQMC benchmark as a second calculation. The unique nuclear-Hamiltonian sample therefore contains
\begin{align}
110+(24-8)=126
\end{align}
points: 110 in $(14,38)$ and eight each in $(14,54)$ and $(38,54)$. No large-space FCIQMC benchmark is included in this sample.

\subsection{Flow diagnostics and sample selection}

For each reverse trajectory, we record the maximum 1- and 2-RDM Hermiticity residuals, one- and ordered two-body trace residuals, contraction residual, contraction-projection correction, natural-occupation violations, and necessary $P$, $Q$, and $G$ violation magnitudes, using the definitions introduced earlier. Tables~\ref{tab:sm_path_algebraic} and \ref{tab:sm_path_physical} report their extrema over the complete 1488-point toy-model sample and the 126 distinct neutron-matter trajectories used for the benchmark comparisons. The neutron-matter statistics thus refer to the same deduplicated sample, not to all 330 trajectories available from the larger density and model-space scan.

All 1488 toy-model flows reach a final off-diagonal norm below $10^{-6}$ MeV. The largest ratio of IMSRG(2) to exact correlation-energy magnitudes is $1.634$, below the acceptance threshold of two. Hermiticity, particle-number traces, and contraction consistency remain close to numerical precision. The largest projection correction, $0.174$, occurs at a strong extended-Hubbard point in the $A=2,L=4$ space, whereas the median trajectory-wise maximum is $2.94\times10^{-6}$. The largest $P$, $Q$, and $G$ violations occur at P3H points in the same space. These positivity tests are not enforced by the contraction projection and do not constitute a complete test of $N$-representability.
Before plotting and fitting Fig.~3 in the main text, we require
\begin{align}
|d_{\rm path}|\geq10^{-6}~{\rm MeV},
\qquad
|d_{\rm MB}|\geq10^{-6}~{\rm MeV},
\end{align}
which is close to our numerical precision.
Both errors are defined per particle. This common display and fit threshold removes 310 toy-model calculations with at least one smaller error and removes none of the 126 nuclear-Hamiltonian calculations. The full unfiltered sample remains in the accompanying dataset.

The displayed and fitted sample therefore contains 1178 toy-model calculations and 126 nuclear-Hamiltonian calculations, totaling 1304 points. Minnesota is included in the nuclear-Hamiltonian group.  
We fit the logarithmic ratio with a Student-$t$ distribution of three degrees of freedom~\cite{Lange1989JASA},
\begin{align}
\log\frac{|d_{\rm MB}|}{|d_{\rm path}|}
\sim
t_{\nu=3}(\mu,\sigma_{\log}),
\end{align}
where $\sigma_{\log}$ is the scale parameter. 
Maximum-likelihood estimation gives
\begin{align}
|\widehat d_{\rm MB}|
=
0.84285\,|d_{\rm path}|,
\qquad
\sigma_{\log}=0.22758.
\end{align}
The fit uses natural logarithms and determines a robust proportional center. The gray line in Fig.~3 in the main text represents this center.
For the same 1304 points, the Pearson correlation between the logarithmic errors is $0.9923$, and the Spearman coefficient is $0.9927$. The correlation extends over more than six orders of magnitude. The remaining scatter is consistent with Hamiltonian-dependent leading coefficients and higher-order corrections in the expansion discussed in the main text.

\begin{table*}[htbp]
\caption{\label{tab:sm_toy_endpoints} Actual coupling endpoints $x_-/x_+$ used to construct the grids. A single positive number denotes a one-sided grid with endpoint $x_+$. A dash denotes a model-space realization not included. All values are expressed in the natural one-body energy unit of the corresponding model. For Kanamori--Hund, $x=J$ at fixed $U=2$ in the $(2,2)$ and $(4,4)$ spaces, whereas $U=2x$ and $J=0.4x$ in the $(2,4)$ and $(4,6)$ spaces.}
\scriptsize
\begin{ruledtabular}
\begin{tabular}{lcccc}
Family &
$A=2,L=2$ &
$A=2,L=4$ &
$A=4,L=4$ &
$A=4,L=6$\\
\hline
Hubbard
& $3.47833$ & $4.31674$ & $2.97528$ & $3.14010$\\
Extended Hubbard
& $-3.14297/3.86959$ & $-3.03118/3.75780$
& $-2.41296/3.25475$ & $-2.35758/2.87926$\\
Ionic Hubbard
& $-6.47705/3.70191$ & $-2.51458/2.68396$
& $-7.12262/3.19886$ & $-1.61327/2.31411$\\
SSH-Hubbard
& $4.72477$ & $2.58233$ & $4.63254$ & $2.96621$\\
Spinless $t$-$V$
& $-14.4/14.4$ & -- & $-1.98997/3.70191$
& $-1.45519/5.61155$\\
Spinless SSH-$V$
& $-14.4/14.4$ & -- & $-2.38853/5.55479$
& $-1.42884/5.39636$\\
Spinless ionic-$V$
& $-14.4/14.4$ & -- & $-1.86416/7.30707$
& $-1.48153/5.68328$\\
Frustrated spinless
& $-14.4/14.4$ & -- & $-1.46240/6.29260$
& $-1.63962/6.41852$\\
LMG
& $0.870332$ & $2.66460$ & $1.50312$ & $1.95770$\\
Unequal pairing
& $6.66150$ & $0.824088$ & $1.27762$ & $0.534503$\\
Anderson impurity
& $2.80385$ & $2.65008$ & $2.10808$ & $2.40106$\\
Periodic Anderson
& $1.98651$ & $1.66770$ & $0.499447$ & $0.505470$\\
Kanamori--Hund
& $0.127248$ & $0.617414$ & $0.462266$ & $0.567157$\\
Gaudin--Yang
& $-5.82300/2.67544$ & $-4.26085/3.64601$
& $-6.29260/3.81369$ & $-2.96621/3.60309$\\
Quadrupole--quadrupole
& $1.35000$ & $1.35000$ & $1.35000$
& $1.05000^\dagger$\\
\end{tabular}
\end{ruledtabular}
\begin{flushleft}
$\dagger$ The $A=4,L=6$ quadrupole--quadrupole calculation undergoes an exact-state crossing between points 13 and 14. Only points $i=1,\ldots,13$, corresponding to $0.0347025\leq\chi\leq0.457005$, remain on the reference-connected target branch and are included.
\end{flushleft}
\end{table*}

\begin{table*}[htbp]
\caption{\label{tab:sm_p3h_points} All P3H coupling pairs $(g,b)$ used in the four model spaces.}
\scriptsize
\begin{ruledtabular}
\begin{tabular}{ccccc}
$i$ &
$A=2,L=2$ &
$A=2,L=4$ &
$A=4,L=4$ &
$A=4,L=6$\\
\hline
1  & $(0.0573567,0.475920)$ & $(-2.16761,-2.38365)$ & $(-1.34831,-0.786214)$ & $(-1.78372,-0.788580)$\\
2  & $(0.0573567,2.96279)$  & $(-2.16761,-1.19183)$ & $(-2.00521,-0.584633)$ & $(-2.34287,-0.517890)$\\
3  & $(0.0573567,6.26750)$  & $(-2.16761,0)$        & $(-2.44538,0)$         & $(-3.92548,0)$\\
4  & $(0.0573567,10.1187)$  & $(-2.16761,0.330071)$ & $(-2.44538,0.313066)$  & $(-3.41617,0.184510)$\\
5  & $(0.0573567,14.4000)$  & $(-2.16761,0.660142)$ & $(-2.00521,0.513429)$  & $(-2.90685,0.314003)$\\
6  & $(0.357068,0.475920)$  & $(-1.08380,-2.38365)$ & $(-0.822138,-0.958798)$& $(-1.28655,-1.13757)$\\
7  & $(0.357068,2.96279)$   & $(-1.08380,-1.19183)$ & $(-1.22269,-0.712967)$ & $(-0.891859,-0.394290)$\\
8  & $(0.357068,6.26750)$   & $(-1.08380,0)$        & $(-1.22269,0)$         & $(-1.96274,0)$\\
9  & $(0.357068,10.1187)$   & $(-1.08380,0.330071)$ & $(-1.22269,0.313066)$  & $(-1.45343,0.157001)$\\
10 & $(0.357068,14.4000)$   & $(-1.08380,0.660142)$ & $(-1.22269,0.626132)$  & $(-1.96274,0.424037)$\\
11 & $(0.755343,0.475920)$  & $(0,-2.38365)$        & $(0,-1.42593)$         & $(0,-1.73545)$\\
12 & $(0.755343,2.96279)$   & $(0,-1.19183)$        & $(0,-0.712967)$        & $(0,-0.867726)$\\
13 & $(0.755343,6.26750)$   & $(0.117216,0.0644041)$& $(0.104442,0.0610861)$ & $(0.0538529,0.0240751)$\\
14 & $(0.619381,8.29734)$   & $(0,0.330071)$        & $(0,0.313066)$         & $(0,0.212019)$\\
15 & $(0.507893,9.68256)$   & $(0,0.660142)$        & $(0,0.626132)$         & $(0,0.424037)$\\
16 & $(1.21948,0.475920)$   & $(0.492602,-1.95460)$ & $(0.535265,-1.42593)$  & $(0.259211,-0.948533)$\\
17 & $(1.21948,2.96279)$    & $(0.600734,-1.19183)$ & $(0.535265,-0.712967)$ & $(0.190814,-0.349124)$\\
18 & $(1.21948,6.26750)$    & $(0.600734,0)$        & $(0.535265,0)$         & $(0.474257,0)$\\
19 & $(0.819979,6.80382)$   & $(0.600734,0.330071)$ & $(0.535265,0.313066)$  & $(0.302923,0.135423)$\\
20 & $(0.672383,7.93970)$   & $(0.492602,0.541317)$ & $(0.438917,0.513429)$  & $(0.374858,0.335163)$\\
21 & $(1.73545,0.475920)$   & $(0.662451,-1.31427)$ & $(1.07053,-1.42593)$   & $(0.381629,-0.698248)$\\
22 & $(1.73545,2.96279)$    & $(0.807867,-0.801385)$& $(1.07053,-0.712967)$  & $(0.542265,-0.496078)$\\
23 & $(1.42307,5.13935)$    & $(1.20147,0)$         & $(1.07053,0)$          & $(0.948515,0)$\\
24 & $(0.956873,5.57913)$   & $(0.985204,0.270658)$ & $(1.07053,0.313066)$   & $(0.806515,0.180278)$\\
25 & $(0.784636,6.51055)$   & $(0.807867,0.443880)$ & $(0.719824,0.421011)$  & $(0.605845,0.270845)$\\
\end{tabular}
\end{ruledtabular}
\end{table*}

\begin{table*}[htbp]
\caption{\label{tab:sm_path_pnm_interactions} Two- and three-nucleon interactions used in the neutron-matter calculations. Each listed NN/NNN pair is a distinct Hamiltonian combination, giving 22 combinations in total. Cutoffs are in MeV, and a dash denotes an NN-only calculation. 
The NN interaction series are EMN~\cite{Entem2017PRC}, NNLO$_{\mathrm{opt}}$~\cite{Ekstrom2013PRL}, NNLO$_{\mathrm{sat}}$~\cite{Ekstrom2015PRC}, Texas~\cite{Hu2026PRL1}, Minnesota~\cite{Thompson1977NPA}, and $\Delta$NNLO$_{\mathrm{GO}}$~\cite{Jiang2020PRC}.
The 3N interactions paired with EMN and NNLO$_{\mathrm{opt}}$ follow Refs.~\cite{Drischler2019PRL} and \cite{Hagen2014PRC}, respectively.
For NNLO$_{\mathrm{sat}}$, Texas, and $\Delta$NNLO$_{\mathrm{GO}}$, the corresponding 3N interactions are specified in the respective interaction references.}
\begin{ruledtabular}
\begin{tabular}{llll}
Two-body force & Three-body force
& Two-body force & Three-body force\\
\hline
LO EMN 500
& --
& N$^2$LO$_{\rm sat}$ 450
& --\\
NLO EMN 500
& --
& N$^2$LO$_{\rm sat}$ 450
& N$^2$LO 450 nonlocal\\
N$^2$LO EMN 450
& --
& N$^3$LO EMN 500
& --\\
N$^2$LO EMN 450
& N$^2$LO 450 nonlocal
& N$^4$LO+ EMN 500
& --\\
N$^2$LO EMN 500
& --
& N$^3$LO Texas 394
& --\\
N$^2$LO EMN 500
& N$^2$LO 500 nonlocal
& N$^3$LO Texas 394
& N$^2$LO 394 nonlocal\\
N$^2$LO EMN 550
& --
& Minnesota
& --\\
N$^2$LO$_{\rm opt}$ 500
& --
& $\Delta$N$^2$LO Go 394
& --\\
N$^2$LO$_{\rm opt}$ 500
& N$^2$LO 400 local
& $\Delta$N$^2$LO Go 394
& $\Delta$N$^2$LO 394 nonlocal\\
N$^2$LO$_{\rm opt}$ 500
& N$^2$LO 500 local
& $\Delta$N$^2$LO Go 450
& --\\
N$^2$LO$_{\rm opt}$ 500
& N$^2$LO 500 nonlocal
& $\Delta$N$^2$LO Go 450
& $\Delta$N$^2$LO 450 nonlocal\\
\end{tabular}
\end{ruledtabular}
\end{table*}

\begin{table*}[htbp]
\caption{\label{tab:sm_path_algebraic} Largest algebraic residuals over the complete TS reverse flows for 1488 toy-model calculations and 126 distinct neutron-matter calculations. The two-body trace is the ordered trace.}
\begin{ruledtabular}
\begin{tabular}{lrrrrrr}
Sample & Number &
$\epsilon_{H}^{(1)}$ &
$\epsilon_{H}^{(2)}$ &
$\epsilon_{{\rm Tr}1}$ &
$\epsilon_{{\rm Tr}2}$ &
$\epsilon_c$\\
\hline
Toy models
& 1488
& $2.17\times10^{-16}$
& $2.48\times10^{-16}$
& $7.55\times10^{-15}$
& $2.66\times10^{-14}$
& $3.62\times10^{-15}$\\
Neutron matter
& 126
& $0$
& $2.63\times10^{-18}$
& $6.39\times10^{-14}$
& $5.91\times10^{-12}$
& $1.39\times10^{-15}$\\
\end{tabular}
\end{ruledtabular}
\end{table*}

\begin{table*}[htbp]
\caption{\label{tab:sm_path_physical} Physical-consistency diagnostics over the same samples as Table~\ref{tab:sm_path_algebraic}. The occupation entry is the maximum distance outside $[0,1]$, and the $P$, $Q$, and $G$ entries are $\max(0,-\lambda_{\min})$. Projection statistics are the maximum and median of the trajectory-wise maximum relative corrections.}
\begin{ruledtabular}
\begin{tabular}{lrrrrrr}
Sample &
occupation &
$\epsilon_{\rm proj}^{\rm max}$ &
$\epsilon_{\rm proj}^{\rm median}$ &
$P$ &
$Q$ &
$G$\\
\hline
Toy models
& $1.18\times10^{-4}$
& $1.74\times10^{-1}$
& $2.94\times10^{-6}$
& $5.47\times10^{-2}$
& $3.12\times10^{-2}$
& $4.58\times10^{-2}$\\
Neutron matter
& $5.99\times10^{-11}$
& $3.07\times10^{-3}$
& $8.88\times10^{-4}$
& $3.33\times10^{-4}$
& $8.80\times10^{-6}$
& $1.68\times10^{-5}$\\
\end{tabular}
\end{ruledtabular}
\end{table*}
\newpage
\bibliographystyle{modified-apsrev4-2.bst}
\bibliography{reference}